\documentclass[letter,11pt]{article}

\usepackage{float}
\usepackage{graphics,graphicx}
\usepackage{hyperref,setspace}
\usepackage{amssymb}
\usepackage{epstopdf}

\begin{document}
\newcommand{\htwo}{{\rm H}_2}
\newcommand{\dal}{\ensuremath{\lsb \Delta \alpha/ \alpha \rsb}}
\newcommand{\dmu}{\ensuremath{\lsb \Delta \mu/\mu \rsb}}
\newcommand{\lsb}{\left[}
\newcommand{\rsb}{\right]}

\input journals.sty

\begin{center}
\includegraphics[width=\textwidth]{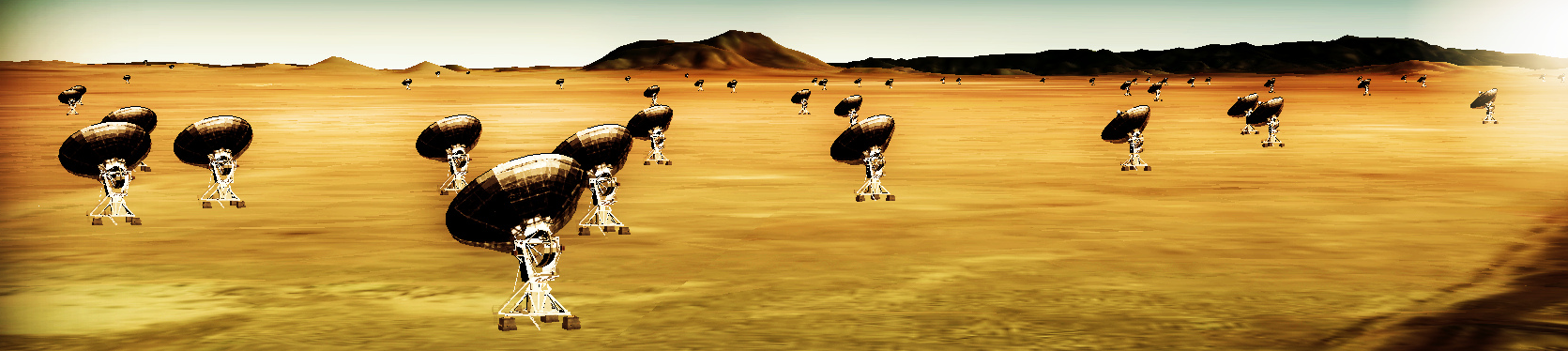}
\end{center}

\begin{center}

{\bf \Large Next Generation Very Large Array Memo No. 9}

\vspace{0.1in}

{\bf \Large Science Working Group 4}

\vspace{0.1in}

{\bf \Large Time Domain, Fundamental Physics, and Cosmology}

\end{center}

\hrule

\vspace{0.7cm}

\noindent Geoffrey C. Bower$^1$,
Paul Demorest$^2$,
James Braatz$^3$,
Avery Broderick$^4$,
Sarah Burke-Spolaor$^2$,
Bryan Butler$^2$,
Tzu-Ching Chang$^5$,
Laura Chomiuk$^6$,
Jim Cordes$^7$,
Jeremy Darling$^8$,
Jean Eilek$^9$,
Gregg Hallinan$^{10}$,
Nissim Kanekar$^{11}$,
Michael Kramer$^{12}$,
Dan Marrone$^{13}$,
Walter Max-Moerbeck$^2$,
Brian Metzger$^{14}$,
Miguel Morales$^{15}$,
Steve Myers$^2$,
Rachel Osten$^{16}$,
Frazer Owen$^2$,
Michael Rupen$^{17}$,
Andrew Siemion$^{18}$
\\[0.2cm]

\vspace{1cm}

\begin{center}
{\bf \large Abstract}\\
\end{center}

We report here on key science topics for the Next Generation Very
Large Array in the areas of time domain, fundamental physics, and
cosmology.  Key science cases considered are pulsars in orbit around
the Galactic Center massive black hole, Sagittarius A*,
electromagnetic counterparts to gravitational waves, and astrometric
cosmology.  These areas all have the potential for ground-breaking and
transformative discovery.  Numerous other topics were discussed during
the preparation of this report and some of those discussions are
summarized here, as well.  There is no doubt that further
investigation of the science case will reveal rich and compelling
opportunities.

\vspace{2mm}

\begin{spacing}{1}
\footnotesize

\noindent $^1$ASIAA, 645 N. A'ohoku Place, Hilo, HI, \\
\noindent $^2$NRAO, PO Box O, Socorro, NM, \\
\noindent $^3$NRAO, 520 Edgemonet Road, Charlottesville, VA, \\
\noindent $^4$Dept. Astronomy, University of Waterloo, Waterloo, ON, Canada \\
\noindent $^5$ASIAA, Roosevelt Rd, Taipei, Taiwan \\
\noindent $^6$Dept. Astronomy, Michgan State Universite, E. Lansing, MI \\
\noindent $^7$Dept. Astronomy, Cornell, Ithica, NY \\
\noindent $^8$CASA, Univ. Colorado, Boulder, CO \\
\noindent $^9$Dept. Physics, New Mexico Tech, Socorro, NM \\
\noindent $^{10}$Dept. Astronomy, Caltech, Pasadena, CA \\
\noindent $^{11}$GMRT, TIFR, Pune, India \\
\noindent $^{12}$MPIfR, auf dem H\"ugel 69, Bonn, Germany \\
\noindent $^{13}$Dept. Astronomy, Univ. Arizona, Tucson, AZ \\
\noindent $^{14}$Dept. of Physics, Columbia Univ, NY, NY \\
\noindent $^{15}$Dept. Physics, Univ. Washington, Seattle, WA \\
\noindent $^{16}$Space Telescope Science Institute, Baltimore, MD \\
\noindent $^{17}$DRAO, Penticton, BC, Canada \\
\noindent $^{18}$UC Berkeley, Berkeley, CA \\
\end{spacing}

\newpage

\tableofcontents

\newpage

\section{Introduction}

Radio astronomy has played a leading role in the areas of fundamental
physics and cosmology.  Examples include the discovery of
gravitational waves with the binary pulsar, the search for the
gravitational wave background with pulsar-timing arrays, the discovery
and characterization of the cosmic microwave background, and
measurements of black hole properties with very long baseline
interferometry.  In addition, the time domain has grown from being a
small component of radio astronomy to a powerful tool for exploring
the Universe. The next generation of radio astronomy instruments,
including the Next Generation Very Large Array (ngVLA), have the
promise to provide discoveries and measurements with similarly
significant impact.  The ngVLA entails ten times the effective
collecting area of the JVLA and ALMA, operating from 1GHz to 115GHz,
with ten times longer baselines (300km) providing mas-resolution, plus
a dense core on km-scales for high surface brightness imaging
\cite{Carilli2015}.

This report is based on discussions held by the Time-Domain, Cosmology
and Physics Scientific Working Group (SWG) in Fall 2014.  It also
reflects additional material that was presented at the January 2015
AAS Workshop on the ngVLA.  The scope of science encompassed by the
SWG was quite broad.  The committee endeavoured to cover the full
breadth in its discussions.  An emphasis was made on identifying
science cases that were transformational rather than ``root-$N$''
improvements on existing cases.  Additional emphasis was placed on
considering the scientific landscape a decade or more in the future,
when ngVLA will come into existence.  Numerous topics, many of them
worthy of further investigation, were discussed by the SWG but are not
presented here for various reasons.  These include strong and weak
gravitational lensing, pulsar timing, fast radio bursts, active
galactic nuclei, cosmic accelerations, the SZ effect, and synoptic
surveys.  There is no doubt that deeper thought on the scientific case
would be valuable.

While many topics were considered of great significance, three topics
were considered among the highest impact:  Galactic Center pulsar
discovery and timing, electromagnetic counterparts to gravitational wave
sources, and astrometry.  Each of these cases has the potential to lead
to transformational discovery.

The technical requirements of the science discussed here is broad.  Some
cases emphasize the longest ngVLA baselines and/or connections to VLB
networks while others are relatively insensitive to angular resolution
and predominantly exploit greater sensitivity.  Broad frequency coverage
was identified as valuable in numerous science cases although there was
not a strict range required.  This SWG is probably unique in
identification of time domain properties as critical.  The science case
does not emphasize fast (i.e., sub-second) transients but many classes
of slower transients will still require high data rates and real-time
processing.

The sections below cover key topics in the areas of time domain,
fundamental physics, and cosmology.

\section{Time Domain}
	\subsection{Radio Gravitational Wave Countparts }

The first direct detection of kHz-frequency gravitational waves (GWs) is
anticipated within the next few years once the ground-based
interferometers LIGO \cite{Abbott+09} and VIRGO \cite{Acernese+09}
achieve their planned ``advanced'' sensitivity.  The most promising
astrophysical GW sources in the frequency range of these detectors are
the inspiral and coalescence of compact object binaries with neutron
star (NS) and/or black hole (BH) constituents, which are expected to be
detected out to distances of hundreds of Mpc.  Although this
accomplishment will stand on its own merits, optimizing the science
returns from a GW detection will require the identification and study of
coincident electromagnetic (EM) counterparts
(e.g.~\cite{MetzgerBerger12}).  This is important for several reasons,
including lifting degeneracies associated with the inferred binary
parameters; reducing the minimum signal-to-noise ratio required for a
confident GW detection; and identifying the merger redshift, thereby
setting the energy scale and allowing an independent measurement of the
Hubble constant or other cosmological parameters.  The potential wealth
of complementary information encoded in the EM signal is likewise
essential to fully unraveling the astrophysical context of the event,
for example an association with specific stellar populations.

Numerical simulations of the merger of neutron stars with other neutron
stars or black holes (e.g.~\cite{Hotokezaka+11}) show that such events
typically eject a few hundredths of a solar mass of matter into space
with high velocities, $\sim 0.1-0.3$ c.  This fast ejecta produces a
blast wave as it is decelerated by its interaction with the surrounding
interstellar medium, powering radio synchrotron emission that could
provide a promising GW counterpart \cite{NakarPiran11}.   This emission
is predicted to peak on a timescale of months to years, at a flux of up
to a few tenths of a millijansky at GHz frequencies.  The black hole
created or newly-fed by the merger is furthermore surrounded by a
massisve torus, the subsequent accretion of which may power a transient
relativistic jet and a short gamma-ray burst (GRB).  The radio afterglow
from the GRB jet may contribute additional bright radio emission.
Importantly, unlike the beamed gamma-ray emission, or the beamed optical
and X-ray afterglow, the radio emission is expected to be relatively
isotropic, making it a more promising counterpart for the majority of
GW-detected events.  Measuring the radio emission from a NS merger would
provide information on the total energy of the ejected matter, its
velocity, and the density of the surrounding medium, the latter
providing information on whether the merger occured within the host
galaxy, or in a lower density medium (e.g. intergalactic space), as
would be expected if the binary was given a natal kick at the time of
formation of the constituent neutron stars.  

One of the biggest uncertainties in predicting the outcome of a NS
merger results from our incomplete knowledge of the equation of state
(EoS) of high density matter.  The recent discovery of massive $\sim
2M_{\odot}$ neutron stars \cite{Demorest+10,Antoniadis+13} indicates
that the EoS is stiffer than predicted by some previous models.  This
opens the possibility that some binary neutron star mergers may result
in the formation of a long-lived neutron star instead of immediately
collapse to a black hole (e.g.~\cite{Siegel+14}).  Such a stable remnant
is formed rapidly rotating, with a period of $P\sim 1$ ms, and
correspondingly large rotational kinetic energy of $E_{\rm rot} \approx
4\pi^{2}I/P^{2}\approx 3\times 10^{52}$ erg, where $I\sim 10^{45}$ g
cm$^{2}$ is the NS moment of inertia.  If the merger remnant also
possesses a moderately strong dipole magnetic field of $B > 10^{13}$ G,
then its electromagnetic dipole spin-down will transfer the rotational
energy to the small quantity of ejecta, accelerating it to
trans-relativistic speeds, $\Gamma_i\simeq E_{\rm rot}/M_{\rm ej}c^{2} >
1$, before the ejecta have been decelerated by the ISM
\cite{MetzgerBower14}.  In this case the radio flux can be orders of
magnitude brighter than in the case of black hole formation, with peak
fluxes of tens or hundreds of millijanksys for a source at typical
distances of LIGO detections.  Detecting such events, combined with
information on the binary mass available from the gravitational wave
signal, could provide stringent constraints on the equation of state of
dense nuclear matter.  

Discovery and localization of EM GW counterparts will require the
ability to survey large areas of sky quickly and sensitively.  EM GW
signals are expected to be faint, short-lived (days), and localized to a
typical accuracy of $> 100$ deg$^2$ \cite{2014ApJ...795..105S}.  Thus, a
fast survey capability is required for radio discovery of the EM
counterpart.  Radio discovery is challenging because of the required
survey speed but faces a much simpler identification problem than in the
optical, where substantial survey capability will exist but the density
of optical transients is very large.  Radio observations will play an
important role in surveying area as well as following-up on candidates
identified through optical or high-energy identification.  Higher
frequency observations are critical for energy calorimetry.  The very
high angular resolution of the ngVLA possibly in coordination with VLBI
networks will be powerful for identification of progenitor systems.
While Advanced LIGO and VIRGO are turning on in the later part of this
decade, uncertain detection rates and source models combined with
current capabilities to detect counterparts indicate that ngVLA
observations could still play an important role in the discovery and
characterization of these sources.

	\subsection{Jetted Tidal Disruption Events}

A rare glimpse into the properties of normally quiescent supermassive
black holes (SMBHs) is afforded when a star passes sufficiently close
that it is torn apart by the SMBH's tidal gravitational field
\cite{Hills75}.  Numerical simulations show that the process of
disruption leaves a significant fraction of the shredded star
gravitationally bound to the SMBH (e.g. Hayasaki et al. 2013).  The
accretion of this stellar debris has long been predicted to power a
thermal flare at optical, UV, and X-ray wavelengths, with over a dozen
such thermal flare candidates now identified (e.g.~\cite{Gezari+13} and
references therein).  

\begin{figure}[htb!]
\begin{center}
\includegraphics[width=1.0\textwidth]{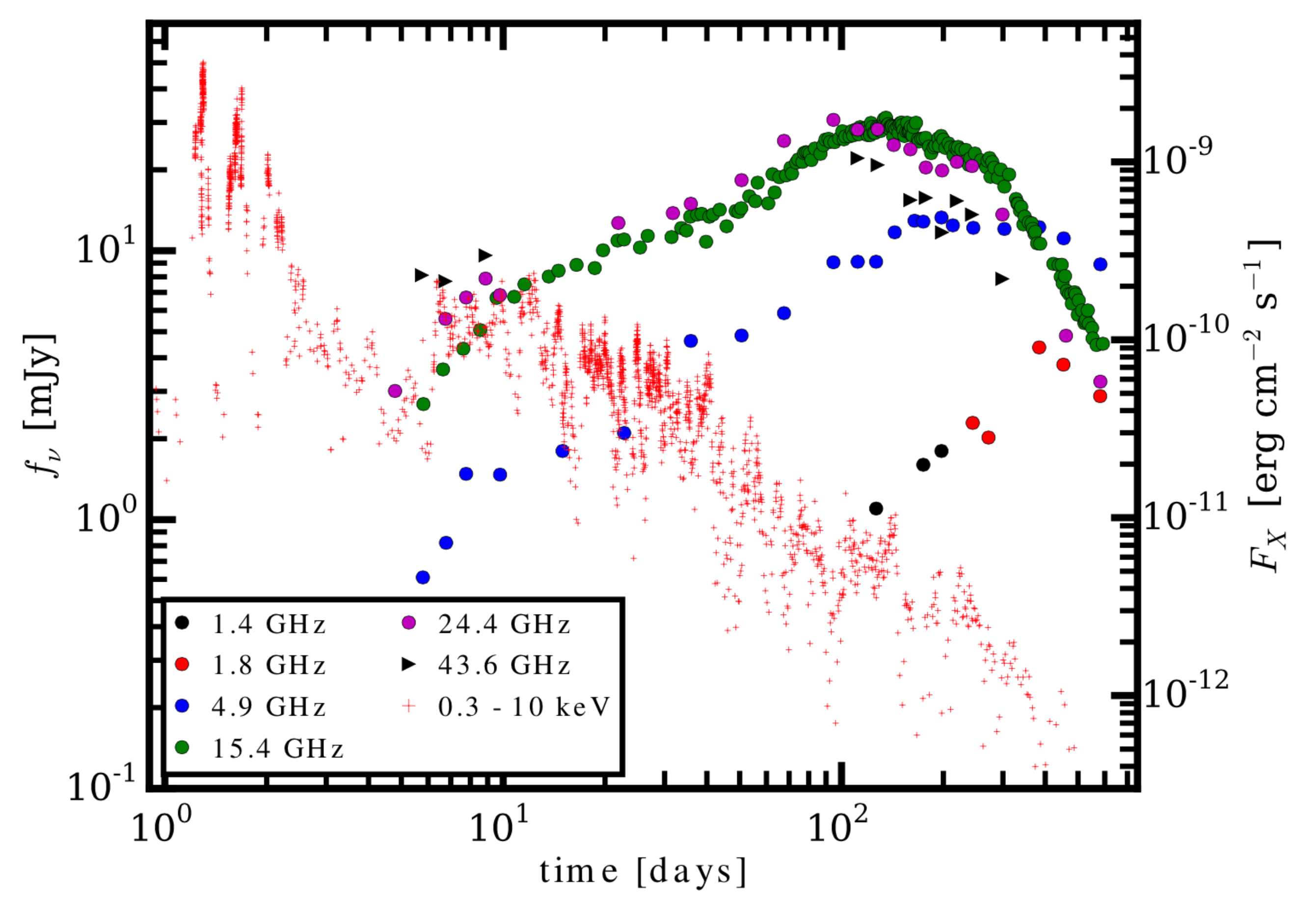}
\end{center}
\label{fig:radio}
\caption{Radio light curves of Swift J1644+57~at observing frequencies
$1.4$, $1.8$, $4.9$, $15.4$, $24.4$ and $43.6$ GHz from \cite{Berger+12}
and \cite{Zauderer+13}.  Also shown are the \emph{Swift} XRT
observations in the $0.3-3$ keV band (red crosses, right axis).  After
several days of peak activity, the X-ray luminosity decreased as a power
law $L_{x} \propto t^{-\alpha}$ in time with $\alpha \sim 5/3$,
consistent with the predicted decline in the fall-back rate $\dot{M}$ of
the disrupted star \cite{Rees88}.  However, the radio emission shows a
broad secondary maximum occurring $\sim 6$ months after trigger, well
after the X-ray maximum.}
\end{figure}

The transient event Swift J1644+57~was characterised by powerful
non-thermal X-ray emission \cite{Bloom+11, Levan+11, Burrows+11,
Zauderer+11}.  The long duration of Swift J1644+57 and a position
coincident with the nucleus of a previously quiescent galaxy led to the
conclusion that it was powered by rapid accretion onto the central SMBH
following a TDE.  The rapid X-ray variability suggested an origin
internal to a jet that was relativistically beaming its radiation along
our line of sight, similar to the blazar geometry of normal active
galactic nuclei \cite{Bloom+11}.  

Swift J1644+57~was also characterised by luminous synchrotron radio
emission, that brightened gradually over the course of several months
\cite{Zauderer+11}, as shown in Figure \ref{fig:radio}.  Unlike the
rapidly varying X-ray emission, the radio emission resulted from the
shock interaction between the TDE jet and the dense external gas
surrounding the SMBH \cite{GianniosMetzger11, Bloom+11}, similar to a GRB
afterglow.  A second jetted TDE, Swift J2058+05, with similar  X-ray and
radio properties to Swift J1644+57~was also reported \cite{Cenko+12}.

Jetted TDEs in principle offer a unique opportunity to witness the birth
of an AGN, thus providing a natural laboratory to study the physics of
jet production across a wide range of mass feeding rates.  Obtaining a
better understanding of these and future jetted TDEs would thus have
far-reaching consequences for topics such as the physics of relativistic
jet formation and super Eddington accretion, the conditions
(e.g.~distribution of accreting or outflowing gas) in nominally
quiescent galactic nuclei, and possibly even the astrophysical origin of
ultra-high energy cosmic rays \cite{FarrarGruzinov09}. 

Despite this promise, many open questions remain regarding how jetted
TDEs fit into the broader phenomenology of TDEs detected at other
wavelengths.  It is currently unclear, for instance, what special
conditions are required to produce a powerful jet in TDEs.
Hydrodynamical models for the evolution of on-axis TDE jets fit to
detailed radio light curves can be used to quantify how similar events
would appear to observers off the jet axis (e.g.~\cite{Mimica+15}).
Such information can be used constrain the presence of off-axis jets in
TDEs detected via their quasi-isotropic thermal emission
\cite{Bower+13,vanVelzen+13} or to assess how jetted TDEs will
contribute to future radio transient surveys \cite{GianniosMetzger11}.  

Dozens of off-axis (`orphan') TDEs similar to Swift J1644+57 could be
discovered by near-term proposed wide-field surveys at GHz frequencies
such as Australian Square Kilometer Array Pathfinder and the Very Large
Array Sky Survey (VLASS) (e.g.~\cite{MetzgerWilliamsBerger15}).  The low
and mid-frequency SKA could detect hundreds of events out to high
redshifts $z > 2$ over the course of a 3 year survey.  The high
sensitivity of the ngVLA would be needed to follow-up and detect these
discovered events at high frequencies, helping to confirm their origin
as TDEs.  The measured redshift distribution of TDEs rate could help
constrain the poorly-understood cosmological evolution of low-mass $<
10^{8}M_{\odot}$ SMBHs.

The high frequency radio light curve of Swift J1644+57 is characterized
by an unexpected flattening or rebrightening on a timescale of a few
months after its initial rise (\cite{Berger+12}; Fig.~\ref{fig:radio}),
with peak fluxes of tens of mJy we obtained at $\sim 10-50$ GHz on
timescales of a couple hundred days.  This behavior contradicted the
expectations of simple one-dimensional jet models that assumed energy is
added to the blast wave in proportion to the observed X-ray luminosity,
and may instead suggest that either the temporal or angular structure of
the jet is more complex than commonly assumed (e.g.~\cite{Berger+12}).
Given the relatively slow evolution of TDE radio light curves, the high
sensitivity of the ngVLA would allow events like J1644+57 to be
monitored for decades or longer.  Better characterizing the emission of
jetted TDEs at high radio frequencies $\sim 10-100$ GHz could be used to
constrain the energy scale and angular structure of the jet, as well as
to trace the gas density in the parsec-scale environment around
previously dormant supermassive black holes.  

As the most sensitive contributor to long baseline observations, the
ngVLA could also play an important role in resolving TDE jets.  The
angular size at 30 GHz of a jetted TDE similar to {\it Swift} J1644+57
near the time of peak radio flux on timescales of a year is expected to
be $\ll 0.1$ mas at the distance of this event ($z = 0.354$;
\cite{Mimica+15}).  However, for a closer off-axis event at $z = 1$, as
could be detected by an upcoming GHz surveys, the jet angular size is
predicted to reach $> 0.5$ mas while its flux still exceeds 1 mJy
(\cite{Mimica+15}; their Fig.~13).  Resolving the dual radio lobes from
such a source would provide invaluable information on the jet angular
structure, such as the possible effects of jet precession due to the
black hole spin \cite{Stone&Loeb11}.


\section{Fundamental Physics}
	\subsection{Pulsars around Sgr A*}

With an average spectral index of $-1.7$, the flux density spectrum of
pulsars is steep. Hence, despite being broadband radio emitters, pulsar
observations at high frequencies require a sensitive telescope.  The
highest frequency that (normal) radio pulsars have been observed at is
134 GHz (Torne et al., in prep.) but the overall sample of
mm-observations is currently very limited \cite{mkt+97,ljk+08}.  But
given the decrease in discovered pulsars for smaller galactocentric
radii (e.g. \cite{lfl+06}) and the overall expectation that interstellar
scattering prevents our view \cite{cl97} to the expected large
population of pulsar in the Galactic Center \cite{wcc+12}, observations
at high radio frequencies is exactly what is needed to reveal this
previously hidden pulsar population. The ngVLA would be
superb instrument to succeed in this task.

The rewards for finding pulsars in the Galactic Center are potentially
huge. While high frequency studies of pulsars are also very interesting
for providing us with clues for identifying the coherent emission
process of pulsars, which still unknown (e.g. \cite{lk05}), the real
motivation is the possibility to study the properties of the central
Black Hole (BH) with a exciting level of detail and precision.

\bigskip
\noindent{\bf Finding pulsars in the Galactic Center.} It has been
demonstrated that pulsars orbiting Sgr A* will be superb probes for
studying the properties of the central supermassive black hole
\cite{wk99,kbc+04,lwk+12}. As shown and discussed further below, it is
even sufficient to find and time a normal, slowly rotating pulsar in a
reasonable orbit, in order to measure the mass of Sgr A* with a
precision of $1M_\odot$(!), to test the cosmic censorship conjecture to
a precision of about 0.1\% and to test the no-hair theorem to a
precision of 1\%. This is possible even with a rather modest timing
precision of $100\mu$s due to the large mass of Sgr A* and the
measurement of relativistic and classical spin-orbit coupling, including
the detection of frame-dragging. Unlike other methods, Liu et al.~also
developed a method that allows us to test for a possible contamination
of the orbital measurements by nearby stars. Given the huge rewards for
finding and timing pulsars in the Galactic centre, various efforts have
been conducted in the past to survey the inner Galaxy and the Galactic
centre in particular (e.g.~\cite{kkl+00,jkl+06,dcl09,mkfr10}). None of
these efforts has been successful, despite the expectation to find more
than 1000 pulsars, including millisecond pulsars (e.g.~\cite{wcc+12}) or
even highly eccentric stellar BH-millisecond pulsar systems \cite{fl11}.
As indicated above, this can be understood in terms of severely
increased interstellar scattering due to the highly turbulent medium.
Scattering leads to pulse broadening that cannot be removed by
instrumental means and that renders the source undetectable as a pulsar,
in particular if the scattering time exceeds a pulse period. The
scattering time, however, decreases as a strong function of frequency
($\propto \nu^{-4}$, see e.g. \cite{lk05}), so that the aforementioned
pulsar searches have been conducted at ever increasing frequencies --
the latest being conducted at around 20 GHz. The difficulty in finding
pulsars at these frequencies is two-fold. We already mentioned that the
flux density is significantly reduced due to the steep spectra of
pulsars. On the other hand, the reduced dispersion delay, which usually
needs to be removed but also acts as a natural discriminator between
real pulsar signals and man-made radio interference, is making the
verification of a real signal difficult. A phased-up ngVLA
that records baseband data would greatly help in both respects. The
large sensitivity would allow us to perform deep searches of the
Galactic Center. The baseband data can be used to study the frequency
structure of the signal in great detail, for instance, by synthesising a
very fine polyphase-filterbank or by searching directly for chirped
signals.

The serendipitous discovery of the magnetar, PSR J1745-2900,
at a separation of 0.1 pc from Sgr A* has led to a new paradigm
for understanding Galactic Center scattering and the pulsar population.
The pulsar shows much less temporal broadening than predicted
based on models for a close scattering screen \cite{2014ApJ...780L...3S},
suggesting that the dominant scattering may occur at much larger
distances \cite{2014ApJ...780L...2B}.  In fact, the pulsar
was detected at frequencies as low as 1 GHz, where the traditional
picture for scattering predicts temporal broadening of 1000 seconds.
Secular evolution of the temporal broadening does suggest that some of
the scattering may be occuring close to Sgr A*.
Reduced temporal broadening, however, does not remove the need for high frequency
observations.  The scattering predicted for the more distant screen still
will lead to timing residuals substantially larger than 100 microseconds for
a typical ordinary pulsar and will make millisecond pulsars undetectable at
frequencies below 10 GHz.  Thus, higher frequency searches and timing
will be an important activity for the ngVLA.

Finding pulsars in the Galactic Center will not only lead to unique
studies of the general relativistic description of black holes (see
below), but we also gain a lot invaluable information about the Galactic
Center region itself: the characteristic age distribution of the
discovered pulsars will give insight into the star formation history;
millisecond pulsars can be used as acceleratormeters to probe the local
gravitational potential; the measured dispersion and scattering measures
(and their variability) would allow us to probe the distribution,
clumpiness and other properties of the central interstellar medium; this
includes measurements of the central magnetic field using Faraday
rotation. Figure 1 shows the scattering and dispersion times for
different frequencies. One can see that it is even possible to detect
millisecond pulsars if the frequency is sufficiently high.

In order to compare the achievable sensitivity to current surveys, we
can compare it to the best GBT searches.  If we assume that the VLA
would have a sensitivity that exceeds that of ALMA significantly,
searches at 43 or even 87 GHz would be comparable to those of the GBT
around 20 GHz, but with the chance of even detecting millisecond
pulsars. At 43 GHz, a significant fraction of the central pulsar
population should be detectable, assuming that our currently known
population is representative.

\bigskip
\noindent{\bf Probing the properties of a super-massive BH.} As pointed
out by \cite{wk99} and \cite{kbc+04}, and described in detail by
\cite{lwk+12}, the discovery of radio pulsars in compact orbits around
Sgr A* would allow an unprecedented and detailed investigation of the
spacetime of this supermassive black hole. As Liu et al. showed, pulsar
timing of a single pulsar orbiting Sgr A*, has the potential to provide
novel tests of general relativity, in particular confirming the cosmic
censorship conjecture and no-hair theorem for rotating black holes.
These experiments can be performed by timing observations with 100
$\mu$s precision, assuming orbits of S-star-like objects as assumed in
the proposal for high-precision optical astrometry observations. It can
also be shown (Liu et al 2012, Wex et al.~in prep.) that for orbital
periods below $\sim0.3$ yr, external perturbations can either be ignored
or a visibility through parts of their orbits only is sufficient. For
full orbits, we expect a $\sim 10^{-3}$ test of the frame dragging and a
$\sim10^{-2}$ test of the no-hair theorem within five years, if Sgr A*
is spinning rapidly. The method is also capable of identifying
perturbations caused by distributed mass around Sgr A*, thus providing
high confidence in these gravity tests. It is worth pointing out that
the analysis is not affected by uncertainties in our knowledge of the
distance to the Galactic center, $R_0$ . A combination of pulsar timing
with the astrometric results of stellar orbits would greatly improve the
measurement precision of $R_0$.

\begin{figure}[htb!]
\begin{center}
\begin{tabular}{cc}
\includegraphics[width=0.47\textwidth]{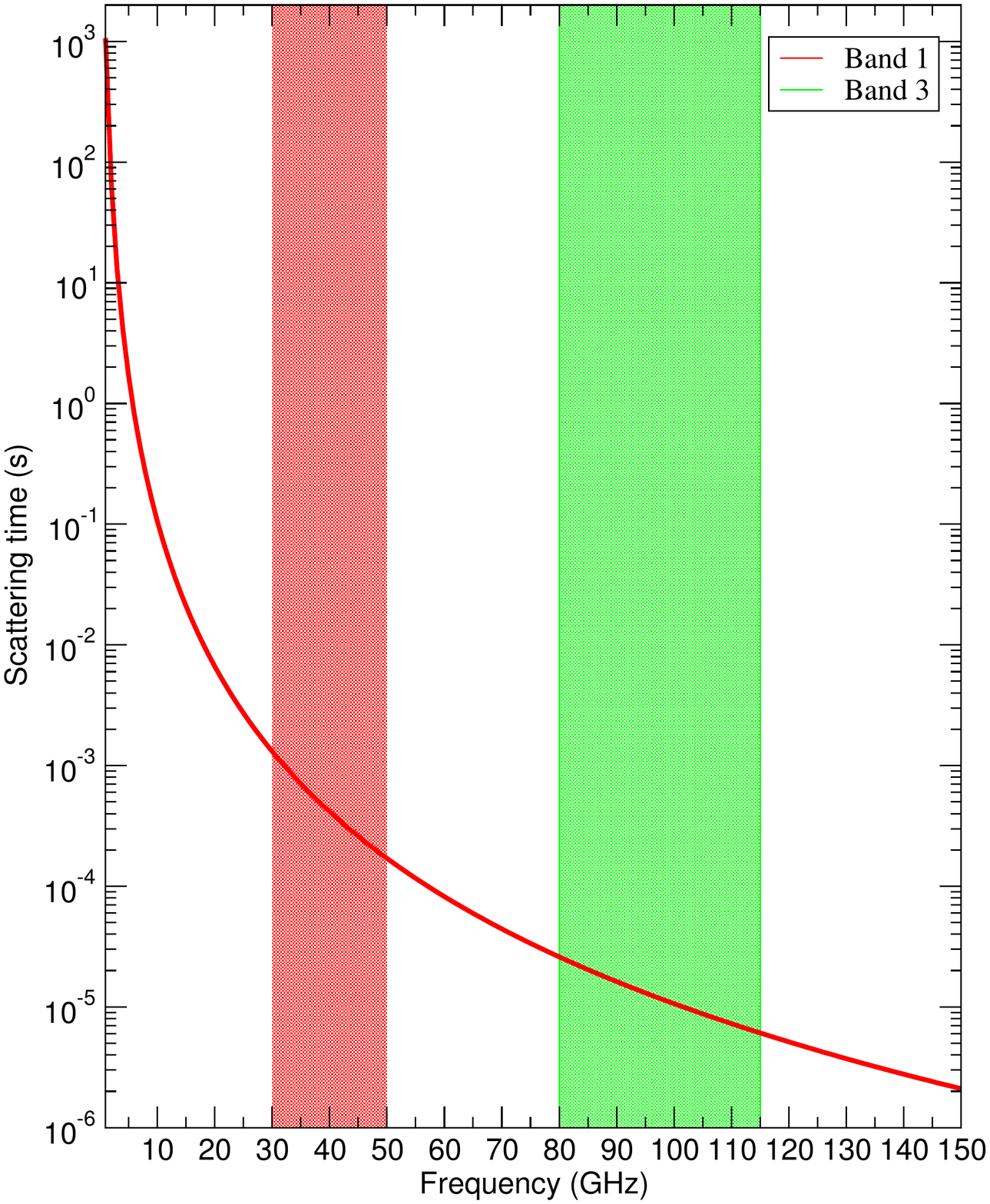} &
\includegraphics[width=0.47\textwidth]{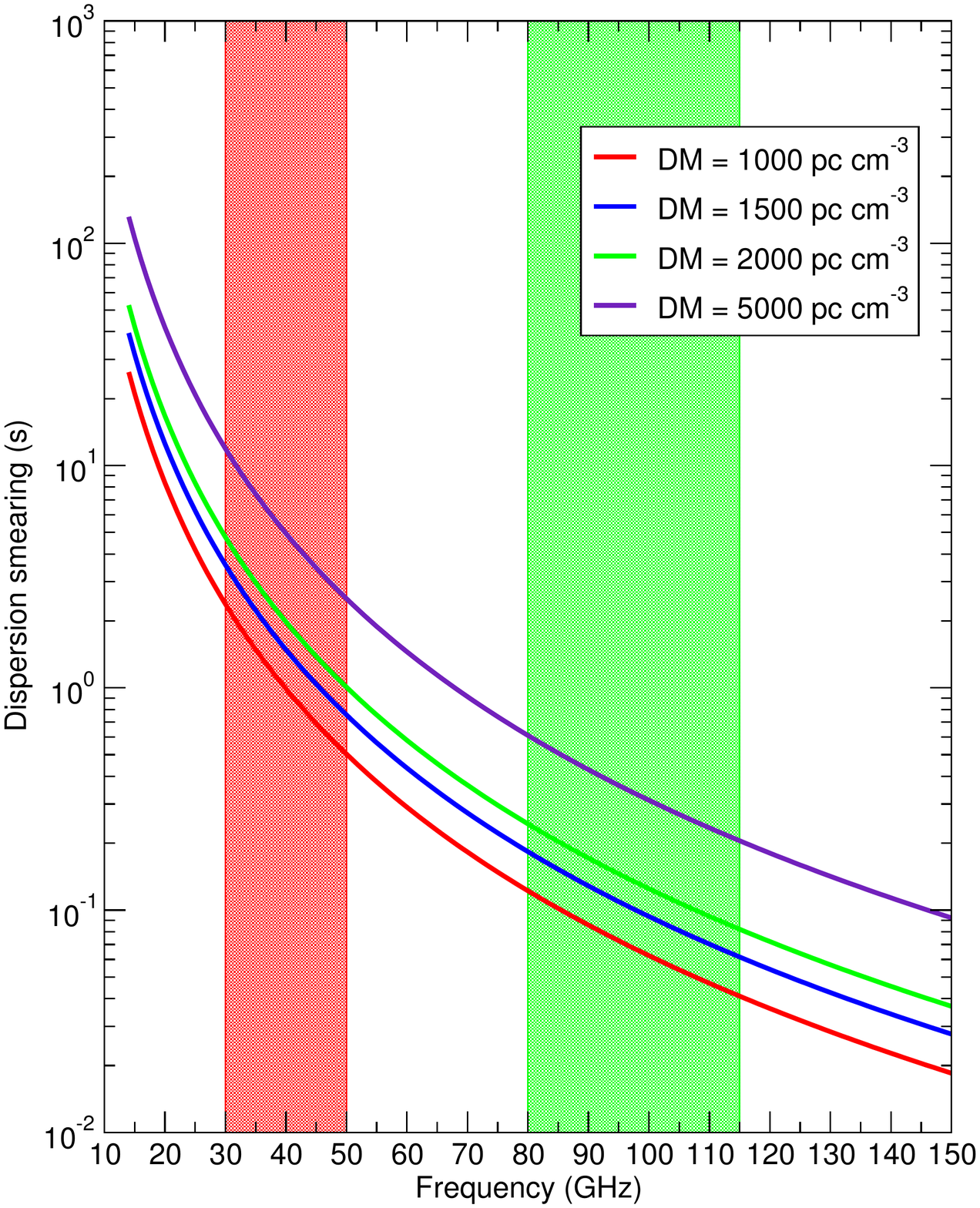} \\
\end{tabular}
\end{center}
\caption{
Scattering time (left) and dispersion smearing (right) as a function of
observing frequency. Possible observing are indicated in green and red.
The scattering time assumes a scattering screen location at 50 pc (see
Cordes \& Lazio 1997).}
\end{figure}

	\subsection{Fundamental Constant Evolution}

Astronomical spectroscopy in redshifted spectral lines has long been
known to provide a probe of changes in the fundamental constants of
physics (e.g. the fine structure constant $\alpha$ and the
proton-electron mass ratio $\mu$, etc.) over timescales of billions of
years. Such temporal evolution is a generic prediction of field theories
that attempt to unify the Standard Model of particle physics and general
relativity (e.g.~\cite{marciano84, damour94}). The exciting possibility
of low-energy tests of such unification theories has inspired a number
of methods to probe fundamental constant evolution on a range of
timescales, ranging from a few years, via laboratory studies, to Gyrs,
with geological and astronomical techniques (see \cite{uzan11} for a
recent review). Most of these methods, both in the laboratory and at
cosmological distances, have been sensitive to changes in the fine
structure constant $\alpha$ (e.g.~\cite{webb99, gould06, rosenband08,
molaro13}).  However, fractional changes in $\mu$ are expected to be far
larger than those in $\alpha$ in most theoretical scenarios, by factors
of $10-500$ (e.g.~\cite{calmet02, langacker02}).  In this discussion, we
will focus on radio spectroscopic probes of changes in $\mu$ over
cosmological timescales.

The astronomical methods to probe fundamental constant evolution are
typically based on comparisons between the redshifts of spectral lines
whose rest frequencies have different dependences on the different
constants. These are so far the only approaches that have found
statistically significant evidence for changes in {\it any} constant:
Murphy et al. \cite{murphy03, murphy04} used the many-multiplet method
with Keck telescope spectra to obtain $\dal = (-5.4 \pm 1.2) \times
10^{-6}$ from 143~absorbers with $\langle z \rangle = 1.75$. This result
has as yet been neither confirmed nor ruled out (e.g.~\cite{srianand07b,
murphy08b, molaro13}. More recently, Webb et al. used the many-multiplet
method to find tentative evidence of possible spatio-temporal evolution
in $\alpha$, combining results from the Keck-HIRES and VLT-UVES
spectrographs \cite{webb11}.  However, these results are all based on
optical spectra, where evidence has recently been found for systematic
effects in the wavelength calibration, due to distortions in the
wavelength scale of echelle spectrographs and the drifts of this scale
with time \cite{griest10, whitmore10, rahmani13}.

For many years, ultraviolet ro-vibrational molecular hydrogen ($\htwo$)
lines provided the only technique to probe changes in $\mu$ on Gyr
timescales \cite{thompson75, varshalovich93, ubachs07}. The resulting
sensitivity to $\dmu$ has been limited by the paucity of redshifted
$\htwo$ absorbers (e.g.~\cite{noterdaeme08}), the low sensitivity of
$\htwo$ lines to changes in $\mu$, and the above systematic effects in
wavelength calibration. Despite a number of detailed recent studies, the
best limits on fractional changes in $\mu$ from this technique are $\dmu
\lesssim 10^{-5}$ ($2\sigma$) at redshifts $0 < z \lesssim 4$
(e.g.~\cite{vanweerdenburg11, rahmani13, bagdonaite15}).

The situation has changed dramatically in recent years with the
development of new techniques using redshifted radio lines from
different molecular species (e.g.~\cite{darling03, chengalur03,
flambaum07b, jansen11, levshakov11, kozlov11}).  While the number of
cosmologically-distant radio molecular absorbers is even smaller than
the number of high-$z$ $\htwo$ absorbers -- just 5 radio systems
\cite{wiklind94, wiklind95, wiklind96, wiklind96b, kanekar05} -- the
high sensitivity of tunneling transition frequencies in ammonia (NH$_3$;
\cite{flambaum07b}) and methanol (CH$_3$OH; \cite{jansen11,levshakov11})
to changes in $\mu$ has resulted in our best present constraints on
changes in any fundamental constant on cosmological timescales. For
example, Kanekar et al. \cite{kanekar11} obtained $\dmu < 2.4 \times
10^{-7}$ ($2\sigma$) from a comparison between the redshifts of CS,
H$_2$CO and NH$_3$ lines from the $z \approx 0.685$ system towards
B0218+357, using the Ku-, Ka-, and Q-band receivers of the Green Bank
Telescope, while Kanekar et al.  (\cite{kanekar15}; see also
\cite{bagdonaite13a,bagdonaite13b}) obtained $\dmu \leq 4 \times
10^{-7}$ ($2\sigma$) from CH$_3$OH lines at $z \approx 0.88582$ towards
PKS1830$-$211, using the Ka- and K-band receivers of the Very Large
Array (VLA). Note that frequency calibration is not an issue at radio
frequencies, with accuracies of $\approx 10$~m/s easily attainable even
with standard calibration techniques.  This is very different from the
situation at optical wavelengths where wavelength errors in today's best
echelle spectrographs are $\approx 0.5-2$~km/s
(e.g.~\cite{griest10,whitmore10,rahmani13}).

Radio studies of fundamental constant evolution today are limited by two
issues. The first is simply the paucity of redshifted radio molecular
absorbers, with only five known systems, all at $z \leq 0.886$. This is
likely to change in the near future with the first ``blind'' surveys for
radio molecular absorption at high redshifts \cite{kanekar14b},
currently being carried out with the VLA and the Australia Telescope
Compact Array (ATCA).  These surveys should yield samples of molecular
absorbers at $z > 1$ that can be followed up in the CH$_3$OH, NH$_3$, OH
and other lines to probe fundamental constant evolution over large
lookback times, $\approx 13$~Gyrs. The second issue is simply the raw
sensitivity of today's telescopes, at the cm- and mm-wave frequencies of
the NH$_3$ and CH$_3$OH lines. This limits both the sensitivity of
studies of fundamental constant evolution using these spectral lines in
systems with detected absorption, as well as the detection of weak
spectral lines whose line frequencies have strong dependences on
$\alpha$ and $\mu$ and hence would yield strong probes of any putative
evolution. For example, the Lambda-doubled CH lines (ground state at
rest frequencies of $\approx 3.3$~GHz) would allow an interesting probe
of changes in both $\alpha$ and $\mu$ \cite{kozlov11}. Similarly,
CH$_3$OH absorption has not so far been detected in the $z = 0.685$
absorber towards B0218+357, and there are multiple CH$_3$OH lines (e.g.
at rest frequencies of $\approx 19$~GHz) that have as yet not been
detected even in the $z = 0.88582$ system towards PKS1830$-$211,
entirely due to sensitivity issues. 

\begin{figure}[htb!]
\centering
\includegraphics[scale=0.4]{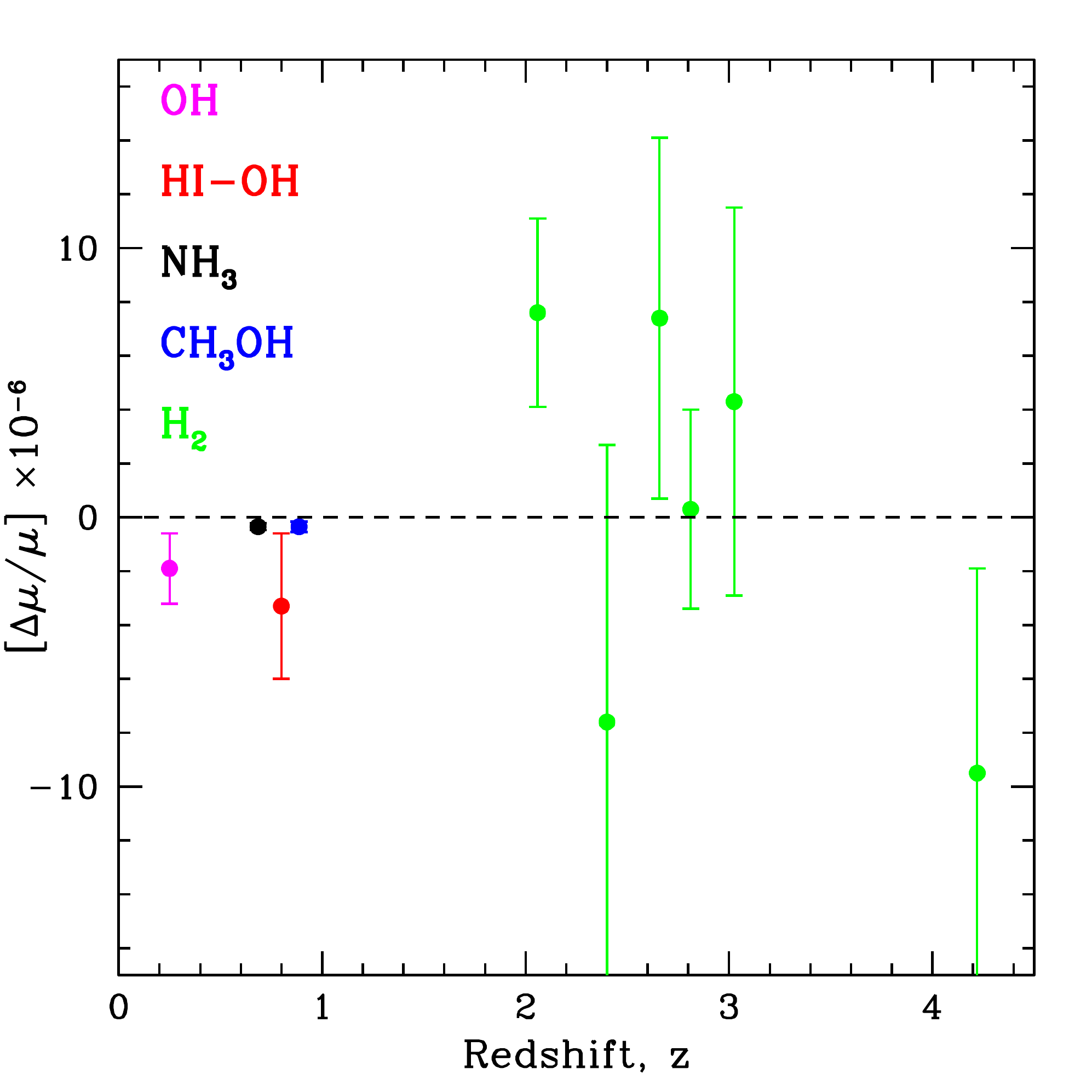}
\caption{
Constraints on changes in $\mu$ from various astronomical methods. The
four results at $z < 1$ are all from radio techniques, based on
comparisons between OH lines \cite{kanekar10}, OH and HI-21cm lines
\cite{kanekar12}, NH$_3$ inversion and rotation lines \cite{kanekar11},
and CH$_3$OH lines \cite{kanekar15}, while the six results at $z > 2$
are all based on ultraviolet $\htwo$ lines \cite{king11,
vanweerdenburg11, wendt12, rahmani13, vasquez14, bagdonaite15}.  It is
clear that the radio results from NH$_3$ and CH$_3$OH lines are more
than an order of magnitude more sensitive than the optical results.
\label{fig:mu}}
\vskip -0.1in
\end{figure}

Fig.~\ref{fig:mu} shows the current state of the art in probes of
changes in $\mu$ on Gyr timescales; it is clear that results from the
NH$_3$ and CH$_3$OH techniques are more than an order of magnitude more
sensitive than those from $\htwo$ lines.  However, the fact that there
are only five radio molecular absorbers currently known, and all at $z
\leq 0.886$, implies that this high sensitivity is currently only
available at $z < 1$, while $\htwo$ lines provide the main technique at
higher redshifts.  Over the next couple of years, deep spectroscopy with
the VLA of the known absorbers towards B0218+357 and PKS1830$-$211 will
allow improvements in the sensitivity to changes in $\mu$ by an order of
magnitude, to $\dmu \approx few \times 10^{-8}$; these studies are
currently under way (e.g. VLA proposal 15A-104). Follow-up VLA and
Atacama Large Millimeter Array spectroscopy of new molecular absorbers
at $z > 1$ detected in the blind VLA surveys should allow sensitivities
of at least $\dal \approx few \times 10^{-7}$.  The limitation in
sensitivity will then be predominantly from the sensitivity of the VLA,
especially since the best NH$_3$ and CH$_3$OH transitions for the
purpose of probing changes in $\mu$ are at cm-wave frequencies, $\approx
12-60$~GHz (e.g.~\cite{flambaum07b, jansen11, levshakov11, kanekar11,
kanekar15}).

The proposed ngVLA would thus have a dramatic impact on studies of
fundamental constant evolution. The increased sensitivity by an order of
magnitude over the VLA would immediately allow an improvement of
sensitivity to fractional changes in $\mu$ by an order of magnitude,
i.e. to $\dmu \approx few \times 10^{-9}$, using the same CH$_3$OH and
NH$_3$ lines in the two known absorbers towards B0218+357 and
PKS1830$-$211. In addition, the increased sensitivity would allow the
detection of multiple CH$_3$OH lines towards B0218+357 as well as the
weaker, but more sensitive, CH$_3$OH lines towards PKS1830$-$211; these
would yield equally sensitive and independent probes of changes in $\mu$
in the two absorbers. Similarly, the ngVLA would allow an improvement in
sensitivity to $\dmu$ by significantly more than an order of magnitude
for the new radio molecular absorbers at $z > 1$, from multiple CH$_3$OH
and NH$_3$ lines.  The ngVLA would also allow the first use of CH lines
to probe changes in $\alpha$ and $\mu$, to $z \lesssim 2.3$. It should
also allow the detection of multiple transitions from species such as
peroxide and methylamine \cite{ilyushin12, kozlov11b}, again providing
independent probes of changes in $\mu$. 

Combining results from different transitions, it should be possible to
achieve sensitivities of $\dmu \approx few \times 10^{-10}$ in the two
absorbers towards B0218+357 and PKS1830$-$211, and of $\approx few
\times 10^{-9}$ on multiple absorbers at $z > 1$ (which will be
identified by the current VLA and ATCA surveys). 
For comparison, next-generation optical facilities (the TMT and the
E-ELT) should achieve sensitivities of $\dmu \approx 10^{-7}$ with
redshifted $\htwo$ lines, over $2 \lesssim z \lesssim 3$. This further
assumes that the wavelength calibration problems of current optical
echelle spectrographs will be fixed in next-generation spectrographs and
that these spectrographs will have high sensitivity in the wavelength
range $3000-4500$~\AA, the wavelengths of the redshifted $\htwo$ lines
from absorbers at $z \approx 2-3$. It is clear that the ngVLA will have
a far higher sensitivity than that of next-generation optical facilities
to changes in $\mu$ with cosmological time.

The techniques described above are based on absorption spectroscopy of
point background sources and hence do not require high angular
resolution. High telescope sensitivity is the critical factor in the
improvement in sensitivity to fractional changes in $\mu$. A wide
frequency coverage is useful, to cover multiple transitions
simultaneously. Extending the frequency coverage to $\approx 100$~GHz
would allow access to high-frequency hydronium transitions
\cite{kozlov11} at $z \gtrsim 2$, well as interesting transitions from
other species (e.g. CH$_3$NH$_2$, CH$_3$OH, etc). Similarly, coverage
down to $\approx 1$~GHz would allow access to the Lambda-doubled OH
lines out to $z \approx 0.7$ as well as the CH lines to $z \le
2.3$. However, the critical science is based on the cm-wave NH$_3$ and
CH$_3$OH lines, and hence requires coverage in the frequency range
$3-50$~GHz, with high spectral resolution and receivers with wide
frequency coverage.

	\subsection{Particle Acceleration and Plasma Physics}

It is generally assumed that relativistic particles (``cosmic rays'') are
energized either in shocks (e.g. supernova remnants) and/or compact
objects (e.g. Active Galactic Nuclei).  However, the data tell us that
particle acceleration also occurs throughout diffuse plasmas (extended
radio galaxies, pulsar wind nebulae, the intracluster medium) where
shocks either should not occur or cannot produce the necessary electron
spectra.  Furthermore, the spectra of these objects are not always
consistent with predictions of shock acceleration theory.

In addition, we know from plasma physics applied to laboratory and space
plasmas that additional particle acceleration mechanisms are possible,
and even common.  One key process is magnetic reconnection (which occurs
when regions of oppositely directed magnetic field are driven together;
resistivity in the consequent current sheet dissipates magnetic energy
and ``reconnects'' field lines). Reconnection is known to occur in earth's
magnetotail, where it is the source for terrestrial aurora, and also in
the sun, where it is the driver for solar flares and Coronal Mass
Ejections.

Good spectral information can discriminate between shock and
reconnection acceleration. Shock acceleration cannot make particle
distributions flatter than $E^{-2}$ \cite{Drury83}; thus the flattest
radio spectrum expected from shock acceleration is $\nu^{-0.5}$.  By
contrast, reconnection acceleration is capable of producing flatter
particle distributions (up to $E^{-1}$; e.g.  \cite{RL92}, \cite{ZH01},
\cite{Guo14}).

We want to know what is going on in more distant objects -- galactic
nebulae, extragalactic systems -- that accelerates relativistic
particles in the extended, diffuse, {\it filamented} regions we observe?
In particular, how important are ``local'' plasma physics processes
(well-understood in space plasmas and the lab) in to more distant
astrophysical sources?  How much have we been missing by only
considering shock acceleration?   We choose two examples to illustrate
how the ngVLA can answer these questions.

\begin{figure}[htb!]
\begin{center}
\includegraphics[width=0.7\textwidth]{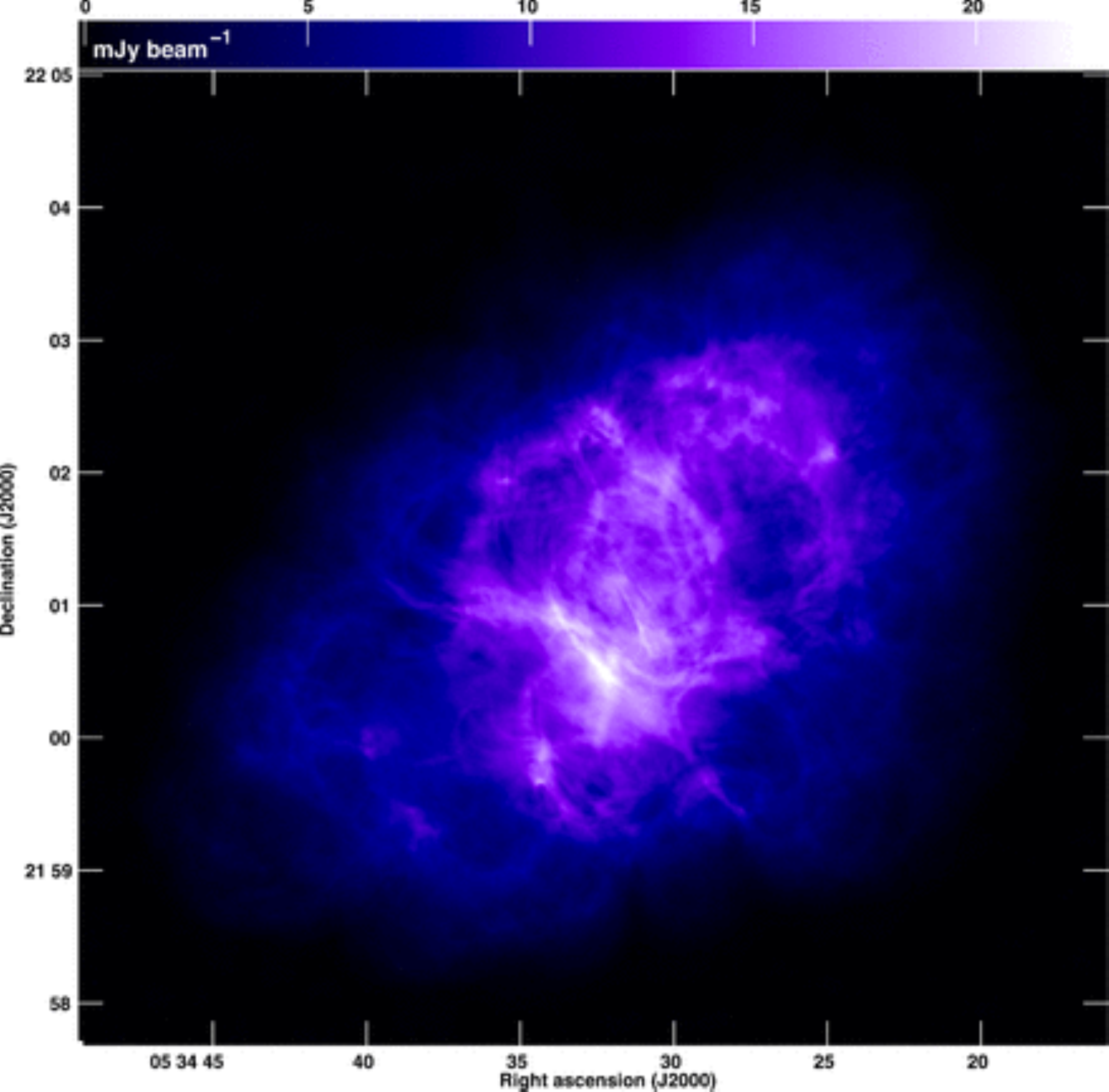}
\end{center}
\caption{\label{fig:CrabNeb} \small
The Crab Nebula at 5 GHz \cite{Biet15}. Current data suggest that
extended particle acceleration is happening throughout the Nebula, {\it
by some means other than shock acceleration} -- possibly magnetic
reconnection.  Are the bright filaments the sites of this acceleration?
Spatially resolved spectral information,  at the highest possible
frequencies, from sensitive ngVLA images will test this hypothesis.
}
\end{figure}

{\it Example:  the Crab Nebula} (Figure \ref{fig:CrabNeb}).  Plasma from
the pulsar passes through a  termination shock and becomes an expanding
wind / synchrotron source which we observe as the Nebula \cite{Biet15}.
While many authors have assumed the shock provides the energization
needed to make the nebula shine, this can't be the full answer (e.g.
\cite{BB14}).  In addition to problems with the likely geometry of the
shock \cite{SS11}, the integrated radio spectrum from the Nebula is too
flat to be consistent with shock acceleration.  Furthermore, while there
are some radio spectral gradients across the nebula \cite{Arendt11}, we
do not see a simple gradient away from the shock.  Instead, there are
flatter spectrum regions throughout the nebula, which suggests
distributed acceleration sites;  but current observations have neither
the resolution nor the frequency sampling to understand the situation.
We want to know what  creates the distributed acceleration sites
throughout the Nebula. Are the synchrotron-bright filaments local shocks
in a turbulent wind?  Or are they local reconnection sites, or the
magnetic flux ropes that are created in a reconnection event?  These
alternatives can be tested with high-resolution spectral imaging, across
the Nebula, to search for small-scale spectral gradients and
correlations with radio-bright filaments.  High frequencies are needed
in order to identify ``pristine'' reconnection regions (the filaments?
elsewhere?), and to learn which process does the accelerating.

\begin{figure}[htb!]
\begin{center}
\includegraphics[width=0.8\textwidth]{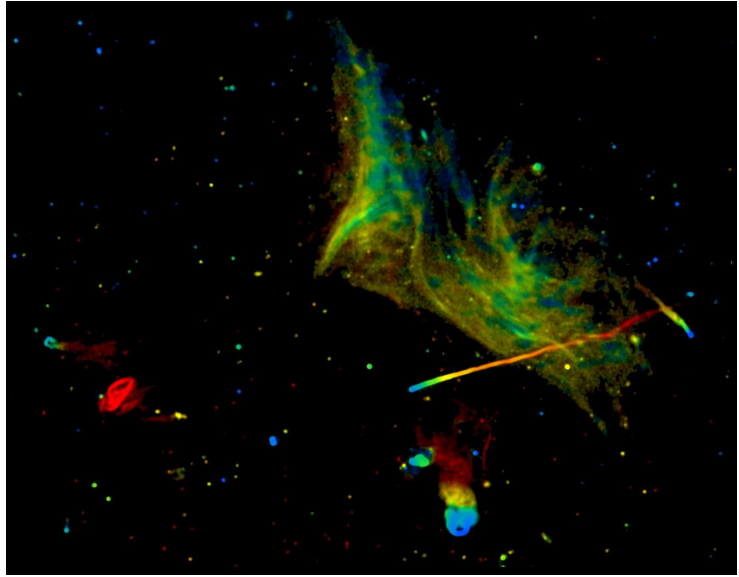}
\end{center}
\caption{\label{fig:A2256} \small
The galaxy cluster Abell 2256 (z = 0.058) at 1.5 GHz \cite{Owen14};  field of view
is $22.'5$. This is a ``true color'' image, shading from
dark blue (spectral index $\alpha \lesssim 0.5$) to red (spectral index $\ge
2$).    The bright, filamented, Mpc-scale structure to the NW is the
``cluster relic''.  We know that in situ particle acceleration is
occurring here, but neither the mechanism nor the nature of the bright
filaments is understood.  Are the filaments post-shock vortices,
reconnection sites, or magnetic pinches similar to the earth's aurora?
Spatially resolved spectra  from ngVLA will be able to discriminate
between these models.
}
\end{figure}

{\it Example:  the Relic in Abell 2256} (Figure \ref{fig:A2256}).
Clusters of galaxies are continually evolving, even at the present
epoch, due to ongoing subcluster mergers.  Many merging clusters have
extended peripheral patches of radio emission (called ``relics'',
\cite{Luigina2012}).  The relics have generally been thought to be
shocks driven by the merger (e.g., \cite{Skill13}), and in some cases
the data support this theory well (e.g. \cite{Stroe14}).  Other cases
are more challenging, however.  Although a large-scale shock has been
invoked to explain the  relic in Abell 2256  \cite{CE06}, the unusual
aspect ratio and strong filamentation of the relic, its high
polarization, and the suggestion of very flat radio spectra in the
relic's outer reaches \cite{Owen14} are hard to explain with a merger
shock. These problems lead one to consider alternative models.  Is the
relic a gigantic reconnection layer?  Are the filaments flux ropes
created in the reconnection event (e.g., \cite{Kagan2013})?  Are they
field-aligned pinches such as seen in the lab \cite{Gregory88} and the
terrestrial aurorae (\cite{Pasch03}), which provide yet another
acceleration mechanism with its own unique spectrum \cite{Lesch97}?
Once again, discriminating between these models requires broad-band,
high-resolution, spectral imaging in order to isolate filament spectra
and compare them to the models.  Establishing that either reconnection
or field-aligned acceleration exists in Abell 2256 will show Mpc-scale,
{\it ordered} magnetic fields  exist in the universe.

\bigskip 
In order to understand the physics of the objects in question we need
good surface brightness imaging, at $\sim$ arcsec resolution, over a
wide field (at least $\sim 10 $ arcmin -- presumably with mosaicing),
and over a broad frequency range.  This is because the critical tests
require high-quality, broadband observations of small features (such as
shocks, filaments, flux ropes, on arcsec scales) which exist within a
broad, diffuse radio-loud region that can be many arcmin across.  We
need good surface brightness imaging because the critical features which
can discriminate between these models are small and bright, but also
embedded in larger, extended emission regions.  We need high frequencies
because spectra are a key test, and a good spectral sampling across a
broad frequency range (at least a decade, ideally more) is needed to
measure spectra robustly.

Since the smallest current configuration of the VLA would produce $\sim
1$ arcsec resolution at 100 GHz, ngVLA must have a very compact
configuration with enough collecting area to yield better surface
brightness sensitivity than the VLA. Furthermore, the large fields of
view required means that ngVLA must work well as a a mosaicing array,
since the individual telescopes will have much smaller primary beams
than many of the sources of interest (e.g. the Crab Neblua, Abell 2256
and the extended emission in M87 \cite{OEK}, \cite{Shi2007}).  This
requirement also argues for smaller array elements, perhaps 12m, in
order to obtain adequate pointing for a mosaic and to reduce the number
of pointings needed. Furthermore some total power mode is needed --
perhaps like ALMA or using a larger single dish -- to sample the largest
spatial frequencies. Given the large number of astrophysically
interesting sources larger than 30 arcsec, this requirement is likely to
be important for many areas of research with ngVLA. For example, the
difficult case of A2256 -- assuming a compact configuration with 3 times
the VLA collecting area, a spectral index of 0.7 and 12m antennas --
the relic could be imaged in the 10-50 GHz band using a 400 pointing
mosaic with typically 20:1 S/N and  3 arcsec resolution in 25 hours.

	\subsection{Star-Planet interactions}

This working group was tasked with envisioning the astronomical
landscape in 10-15 years and then asked to consider the kinds of science
questions expected to be topical at that time for which a next
generation VLA could provide ground-breaking science.  A major effort is
underway to find and characterize extrasolar planets, including the
environment in which those planets exist: radio observations can play an
important role in this effort, through characterization of star-planet
interactions.  Recent work on star-planet interactions have focussed on
``hot Jupiters'', expected to have more complex interactions with their
host star due to the extreme proximity. These systems are being used as
test cases for the kinds of interactions which might be expected from
Earth-like planets. It is technologically easier to find Earth-like
planets around low-mass stars (rather than solar-like ones), and these
are the types of systems which will be searched for in the nearby
universe. Space missions like K2 and TESS, and ground-based searches
like the MEarth project all have the goal to find nearby transiting
Earth-sized planets so that near-future missions like the James Webb
Space Telescope and the next generation of Giant Segmented Mirror
Telescopes (E-ELT, TMT, GMT) can perform follow-up characterization of
the exoplanet atmosphere. Radio observations provide a unique way to
characterize the nature of the accelerated electron population near the
star, as well as provide the only method for direct constraints on cool
stellar mass loss, important for understanding the radiation and
particle environment in which close-in exoplanets are situated.

\begin{figure}[htbp]
\includegraphics[scale=0.43]{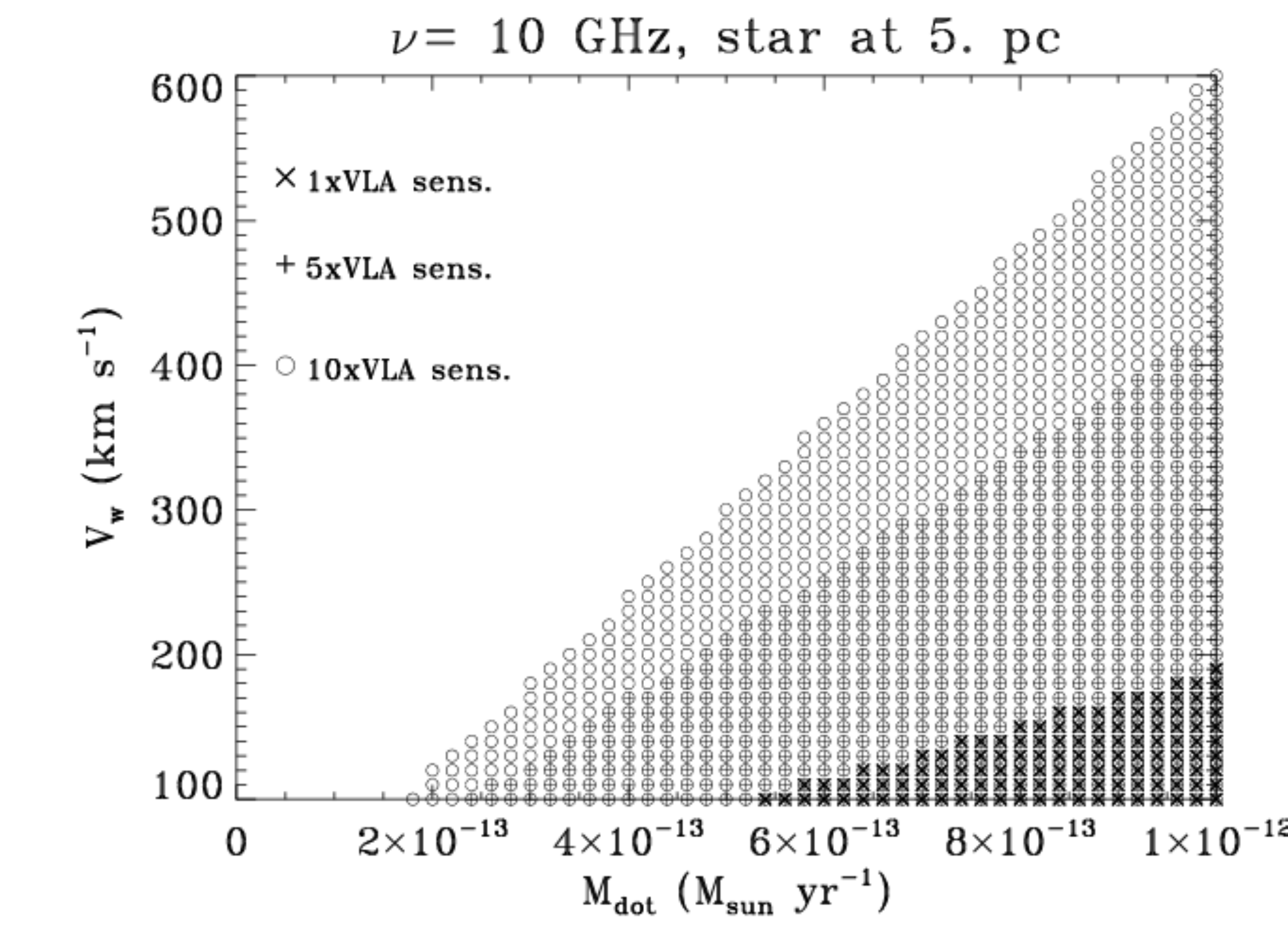}
\includegraphics[scale=0.43]{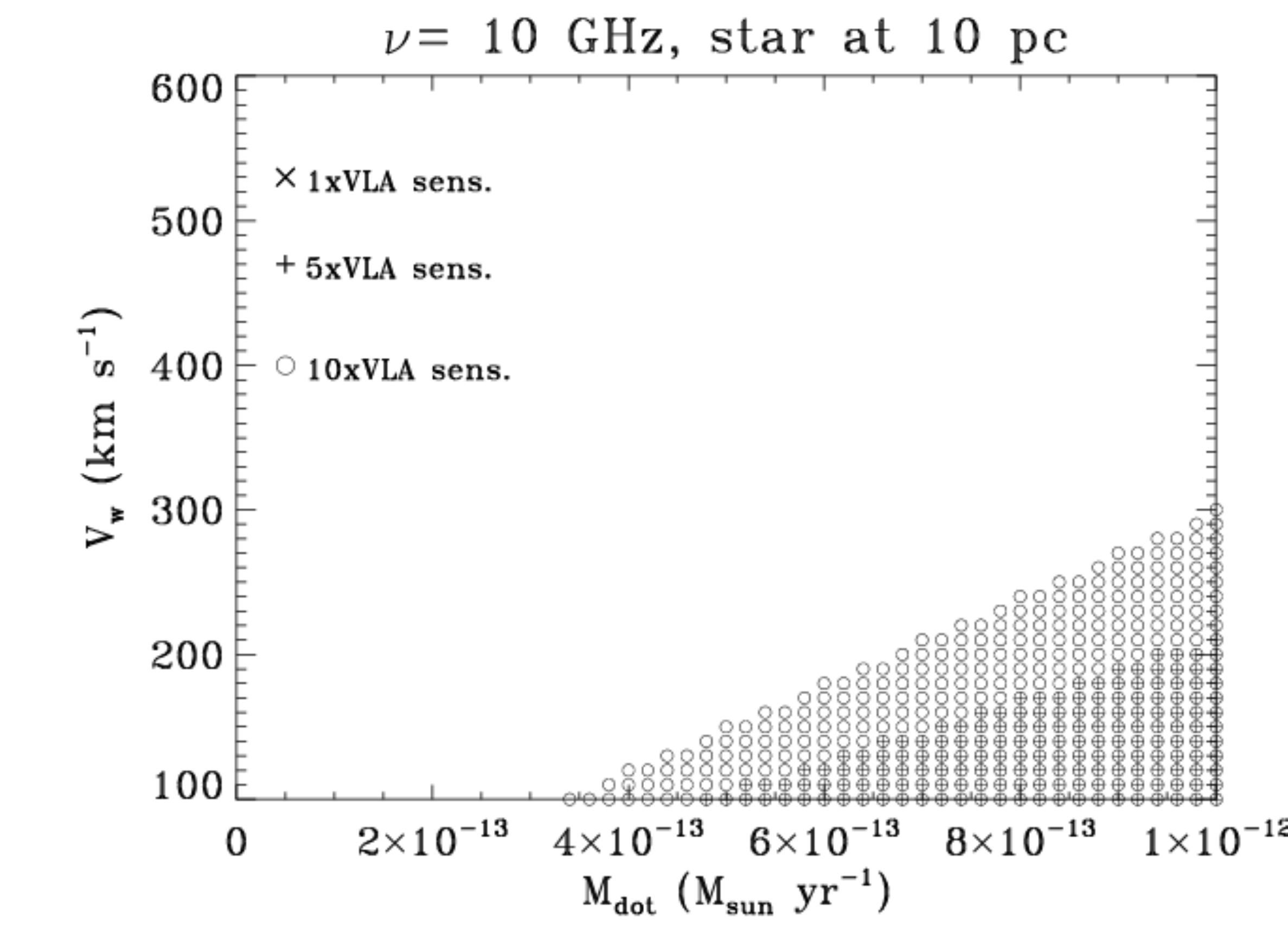}
\caption{
Sensitivity to detecting radio emission from an ionized stellar wind, as
a function of the wind parameters $v_{w}$ and $\dot M$, for emission at
10 GHz, and assuming the star is at 5 pc \textbf{(left)} and 10 pc
\textbf{(right)}. Three different scenarios are employed: the current
VLA sensitivity, scaled up by a factor of 5 (100 nJy rms in 10 hrs) and
scaled up by a factor of 10 (50 nJy rms in 10 hrs).  \label{fig:mdot}}
\end{figure}

Star-planet interactions are important tools to learn about the ecology
set up by magnetically active stars and their close-in planetary
companions. Magnetic reconnection is an inherently non-linear process,
motivating observations to make headway, and while there are detailed
observations of the Sun, its heliosphere may behave completely
differently from that of a magnetically active star with a close-in
exoplanet.  Stellar radio astronomy excels at constraining the
conditions present in outer stellar atmospheres and the stellar near
environments: for magnetically active stars, this is typically
nonthermal coronal emission \cite{gudel2002}, although all stars have
some amount of mass loss associated with them.  Cool stellar mass loss
is characterized by an ionized stellar wind, whose radio flux can have a
$\nu^{0.6}$ or $\nu^{-0.1}$ dependence if in the optically thick or thin
regime, respectively \cite{wrightbarlow1975, panagia1975}. Mass loss in
the cool half of the HR diagram, along/near the main sequence, has been
notoriously difficult to detect, due in part to the much lower values of
mass loss here compared to other stellar environments (upwards of
10$^{-5}$ solar masses per year for massive stars, compared with
$2\times$ 10$^{-14}$ solar masses per year for the Sun). Most efforts to
detect cool stellar mass loss to date rely on indirect methods; i.e.
inferring stellar mass loss by examining absorption signatures of an
astrosphere (e.g.~\cite{wood2005}).  A direct measurement of stellar
mass loss through its radio signature would be a significant leap
forward not only for understanding the plasma physics of the stars
themselves, but also for understanding what kind of environment those
stars create.  Previous attempts at a direct detection of cool stellar
mass loss via radio emission have led to upper limits typically two to
three orders of magnitude higher than the Sun's present-day mass loss,
while indirect methods find  evidence for mass loss rates comparable to
or slightly higher than the Sun's present day mass loss rate (up to
$\sim$80 times solar $\dot M$.) Figure~\ref{fig:mdot} shows the
sensitivity to detecting such a feeble ionized stellar wind, for a radio
telescope with the sensitivity of the VLA, and then 5 and 10 times that
sensitivity, for a star at 5 and 10 pc.  Only for the nearest stars
($\leq$5 pc) and with enhanced, low velocity mass loss are conditions
favorable for the present VLA, and enhanced sensitivities are needed to
make direct detections over a wider range of stellar parameters. The
flat/increasing frequency dependence argues for sensitivity at higher
frequencies to probe the transition where this changeover occurs.  There
are already hints at evaporating planetary atmospheres due to radiation
effects of host stars \cite{vm2003}, with some evidence for the
particle influence of the stellar wind on planetary atmosphere
evaporation \cite{LdE2012}. These inferences can be tested directly
with radio frequency measurements as described above.  

Another aspect of star-planet interactions which can be probed with
radio observations is the magnetospheric interactions of a magnetized
close-in planet with its host star. Similar to the idea that the
proximity of two stellar magnetospheres due to close passage can cause
periodic episodes of magnetic reconnection (as in \cite{massi2006}),
recent results \cite{spi2014} have suggested a triggering mechanism
for regularly recurring stellar flares on stars hosting close-in
exoplanets (hot Jupiters).  To date there have not been any experiments
to test this at radio wavelengths, but one can immediately see the
advantages of using radio wavelength observations, as the time-dependent
response of radio emission reveals the changing nature of the magnetic
field strength and number and distribution of accelerated particles.  As
stellar radio emission typically has a peak frequency near 10 GHz, a
wide bandwidth system spanning this range can probe the optically thick
and thin conditions, diagnose the changing conditions during the course
of a magnetic reconnection flare associated with close passage of an
exoplanet, and deduce the nature of the accelerated particle population
through measurements of spectral indices from confirmed optically thin
emission. This would provide constrains on the accelerated particle
population of close-in exoplanets unavailable from any other
observational method; such a constraint is necessary to perform detailed
modelling of the atmospheres of such exoplanets, due to the influence of
accelerated particles in affecting the chemical reactions in terrestrial
planet atmospheres \cite{jackman1990}.  

\section{Cosmology}

	\subsection{Megamasers}

Exquisite observations of the CMB by WMAP and Planck determine the
angular-size distance to the surface of last scattering at z $\sim$
1100, constrain the geometry of the universe to be very nearly flat, and
set a basic framework for cosmology.  The CMB, however, does not
uniquely determine all fundamental cosmological parameters on its own.
Observations at z $\sim$ 0, when dark energy is dominant, provide
complementary data that constrain critical parameters, including the
dark energy equation of state, the geometry of the universe, and the
number of families of relativistic particles.  CMB observations can {\it
predict} basic cosmological parameters, including $H_0$, but only in the
context of a specific cosmological model.  Comparing CMB predictions to
astrophysical measurements of $H_0$ therefore makes a powerful test of
cosmological models.

In the context of the standard model of cosmology, i.e. a geometrically
flat $\Lambda$CDM universe, Planck measurements predict $H_0$ = 67.8
$\pm$ 0.9 km s$^{-1}$ Mpc$^{-1}$ \cite{2015arXiv150201589P}.  This
result is in tension with recent astrophysical measurements of $H_0$
based on standard candle observations: $H_0$ = 73.8 $\pm$ 2.4 km
s$^{-1}$ Mpc$^{-1}$ \cite{2011ApJ...730..119R} and $H_0$ = 74.3 $\pm$
2.6 km s$^{-1}$ Mpc$^{-1}$ \cite{2012ApJ...758...24F}.  The implications
of this disagreement are of fundamental importance to our understanding
of cosmology, so independent measurements of $H_0$ with unrelated
uncertainties are critical to clarify the situation.

Observations of circumnuclear water vapor megamasers in AGN accretion
disks can be used to measure direct, geometric distances to their host
galaxies, a technique pioneered with observations of NGC
4258 \cite{1999Natur.400..539H}.  At only 7.6 Mpc, though, NGC 4258 is
too close to measure $H_0$ directly.  However, similar megamasers well
into the Hubble flow at D $\sim$ 50 - 200 Mpc are being discovered and
studied to measure distances to their host galaxies.  With these, the
megamaser technique provides $H_0$ in one step, independent of standard
candles and distance ladders.  

Over the past decade, astronomers have made an intensive effort to
discover and measure H$_2$O megamasers, spearheaded mainly by the
Megamaser Cosmology Project (MCP).  Measuring $H_0$ with this technique
requires three types of observations.  First is a large survey to
identify the rare, edge-on disk megamasers suitable for distance
measurments.  Second is sensitive spectral monitoring of those disk
megamasers to measure secular drifts in maser lines, indicative of the
centripetal accelerations of maser clouds as they orbit the central
black hole.  And third is sensitive VLBI observations to map the maser
features and determine the rotation structure and angular size of the
disk.  The map and acceleration measurements together constrain a model
of a warped disk and determine the distance to the host galaxy.  The
measurement is independent of standard candles and provides $H_0$ in a
single step.

A second important goal in discovering and mapping megamasers is to
measure ``gold standard'' masses of supermassive black holes (SMBH) by
tracing the Keplerian rotation curves of megamaser disks only tenths of
a pc from the nuclei and well within the SMBH sphere of influence.
About 20 SMBH masses have been measured, to date, using this technique,
most with $<$ 10\% uncertainties \cite{2011ApJ...727...20K}.  These
measurements demonstrate a breakdown of the M-$\sigma$ relation at the
low-mass end, implying that any feedback that controls the apparent
co-evolution of SMBHs and their elliptical host galaxies has not yet
taken hold in spiral galaxies \cite{2010ApJ...721...26G}.  Larger samples
of SMBH mass measurements are required to trace the shape of the
M-$\sigma$ correlation in detail and constrain galaxy and SMBH growth
models.

\begin{figure}[htb!]
\begin{center}
\includegraphics[width=1.0\textwidth]{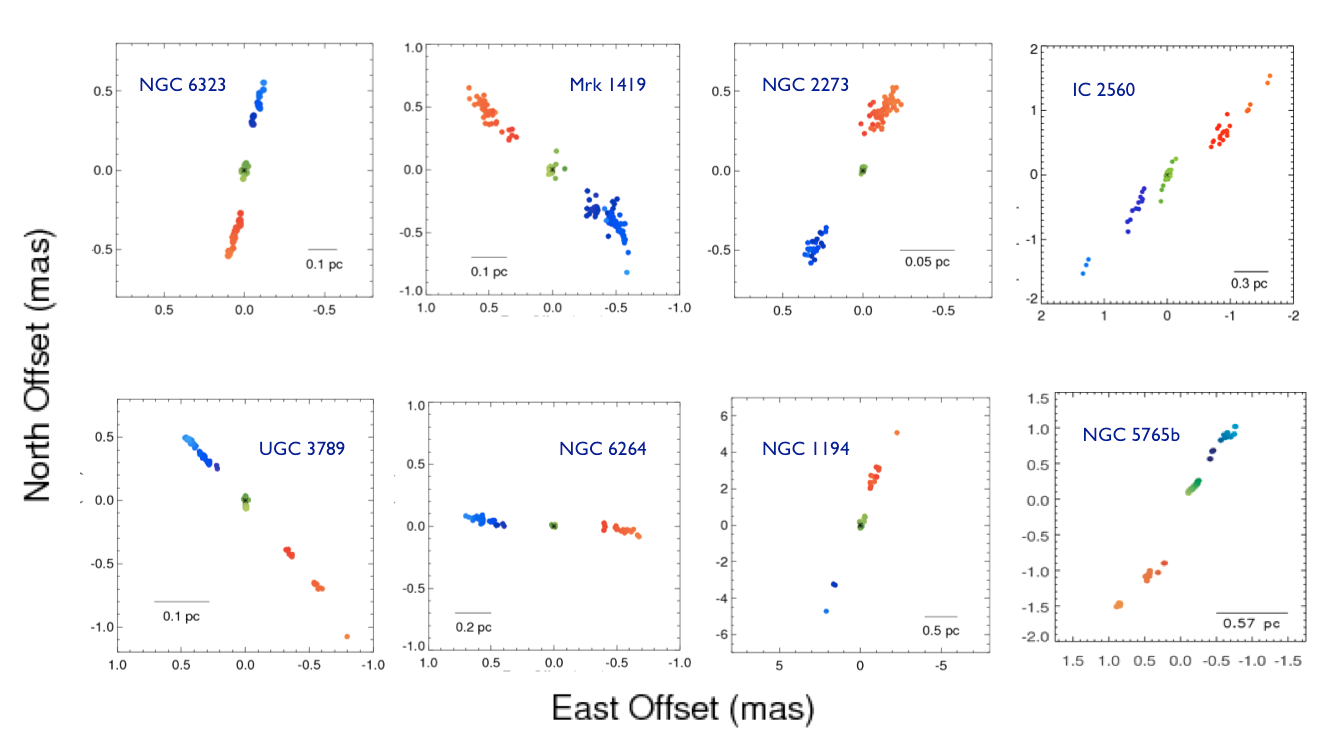}
\end{center}
\label{fig:masers}
\caption{VLBI maps of eight $H_2O$ megamasers showing edge-on disks.
These are among the masers currently being studied to measure H$_0$
directly, using the geometric megamaser technique.  Colors represent
the line-of-sight velocities of maser clouds.  These masers all
demonstrate Keplerian rotation.}
\end{figure}

Of the $\sim$ 30 megamaser systems that have been identified as edge-on
disks, ten of them are bright enough and otherwise appropriate for
measuring distances.  Figure \ref{fig:masers} shows VLBI maps of eight
disk megamasers.  Discovering these rare megamasers has required a
decade of surveys looking at over 3000 Seyfert 2 galaxies with the GBT.
After it is identified as a distance candidate, each megamaser requires
a sensitive VLBI map and about two years of monitoring to determine its
distance.  So far, distances to three have been published: UGC
3789 \cite{2013ApJ...767..154R}, NGC 6264 \cite{2013ApJ...767..155K}, and
NGC 6323 \cite{2015ApJ...800...26K}.  Including these published results
as well as new work in progress, the megamaser technique determines
$H_0$ = 70.4 $\pm$ 3.6 km s$^{-1}$ Mpc$^{-1}$, a 5.1\% result.  This
measurement is intermediate between the Planck prediction and the best
measurements based on standard candles.  In about a year, the MCP will
finish measurements and analysis on the remaining disk galaxies, and
expects to acheive a $\sim$ 4\% measurement.  The MCP uses the Green
Bank Telescope for surveys and spectral line monitoring, and the High
Sensitivity Array (VLBA + GBT + phased VLA + Effelsberg) for VLBI
mapping.

The long-term goal of the observational cosmology community is to reach
a $\sim$ 1\% measurement of $H_0$ that is robust to the different
measurement techniques.  To approach a 1\% measurement with megamasers,
we could measure, for example, $\sim$ 10\% distances to each of $\sim$
100 megamaser galaxies.  Such a measurement requires significantly
better spectral line sensitivity at 22 GHz than we can achieve with
current instruments.  In addition, to improve sensitivity of VLBI
imaging, approximately 20\% of the collecting area should cover long
($\sim$ 5000 km) baselines.  Individual maser spots are unresolved, so
uv-coverage requirements are modest.  For example, the long baselines
could be achieved with $\sim$ 5 100-m class apertures.  However, long
baselines in both the N-S and the E-W directions are important.  The
recording and correlation capabilities of the new telescope are also
modest, but should permit imaging 1 GHz bandwidths contiguously with at
least 25 kHz spectral resolution.

	\subsection{Intensity Mapping}

The origin and evolution of structure in the Universe is one of the
major challenges of observational astronomy.  How and when did the first
stars and galaxies form?  How does baryonic structure trace the
underlying dark matter? A multi-wavelength, multi-tool approach is
necessary to provide the complete story or the evolution of structure in
the Universe.  Intensity mapping is a critical piece of this mosaic.
The technique relies on the ability to detect many objects at once
through their integrated emission rather than direct detection of
individual objects.  In particular, this provides a window on lower
luminosity objects that cannot be detected individually but that
collectively drive important processes.  

Intensity mapping experiments that are designed to measure the molecular
content of the early universe will provide unique constraints on the
history of metal enrichment, interstellar chemistry, galaxy and stellar
mass growth, and the structure of the reionization process. This concept
fills an otherwise unaddressed gap in our understanding of the growth of
galaxies and  complements many major facilities and initiatives of the
coming decade.  ALMA and EVLA will study the most massive molecular gas
reservoirs at $z>2$, while intensity mapping experiments provide
essential context for these rare giants by measuring the emission of the
bulk population of galaxies.  Similarly, major optical and infrared
facilities, including JWST, will reveal the formed stars and ionizing
radiation in the EoR, but the natal material that fuels this star
formation can only be observed in molecular lines. The radio arrays
searching for EoR 21cm emission neutral hydrogen
\cite{2010Natur.468..796B, 2013A&A...550A.136Y, 2014ApJ...788..106P}
will trace the ionizing bubbles around the galaxies seen with molecular
intensity mapping; ultimately, the cross-correlation of the molecular
and neutral atomic gas signals will provide a direct view of the growth
of ionization bubbles and an important, systematically cleaner,
validation of 21cm power spectra \cite{2011ApJ...741...70L}.  Finally,
the high signal-to-noise power spectrum of molecular material will show
the growth of galaxies over the third billion years of cosmic history.
Cross-correlation with existing and planned optical surveys such as the
Subaru Hyper Suprime Camera (HSC) and Large Synoptic Survey Telescope
(LSST) surveys will explore the gas content, metal enrichment, and
evolution of individual object classes.

Molecular gas is a poorly understood but vital component of galaxy
evolution.  This phase of the interstellar medium (ISM) dominates the
baryonic mass of early galaxies but is extremely difficult to observe.
Rotational transitions of carbon monoxide (CO) are the primary probe of
molecular gas. These transitions occur at integer multiples of the
$J=1\rightarrow 0$ transition at 115 GHz; observed frequencies are
shifted by the cosmological redshift.  With the most sensitive arrays
available it is possible to detect the molecular ISM out to redshifts as
high as $z=6.4$ \cite{2005ARA&A..43..677S,2011AJ....142..101W,
2013Natur.496..329R,2013Natur.495..344V}.  These detections demonstrate
the rapid chemical enrichment of early galaxies, but they probe massive
objects that are not representative of the overall population of
galaxies.  The CO-detected galaxies are typically the most massive and
most metal rich objects of their era, significantly more luminous and
massive than a Milky-Way type galaxy.  ALMA and EVLA observations will
transform our understanding of the molecular gas population down to
Milky-Way like objects with molecular mass limits of $\sim 10^{9}
M_\odot$, primarily through follow-up of galaxies discovered through
other surveys \cite{2011MNRAS.412.1913I,2013IAUS..292..184W}.  In the
study of reionization and galaxy formation, however, it is the low-mass
galaxies, rather than the massive ones, that must be examined because of
their very large number \cite{2012ApJ...752L...5B,2013ApJ...768...71R}.

\begin{figure}[htb!]
\begin{center}
\includegraphics[width=0.9\textwidth]{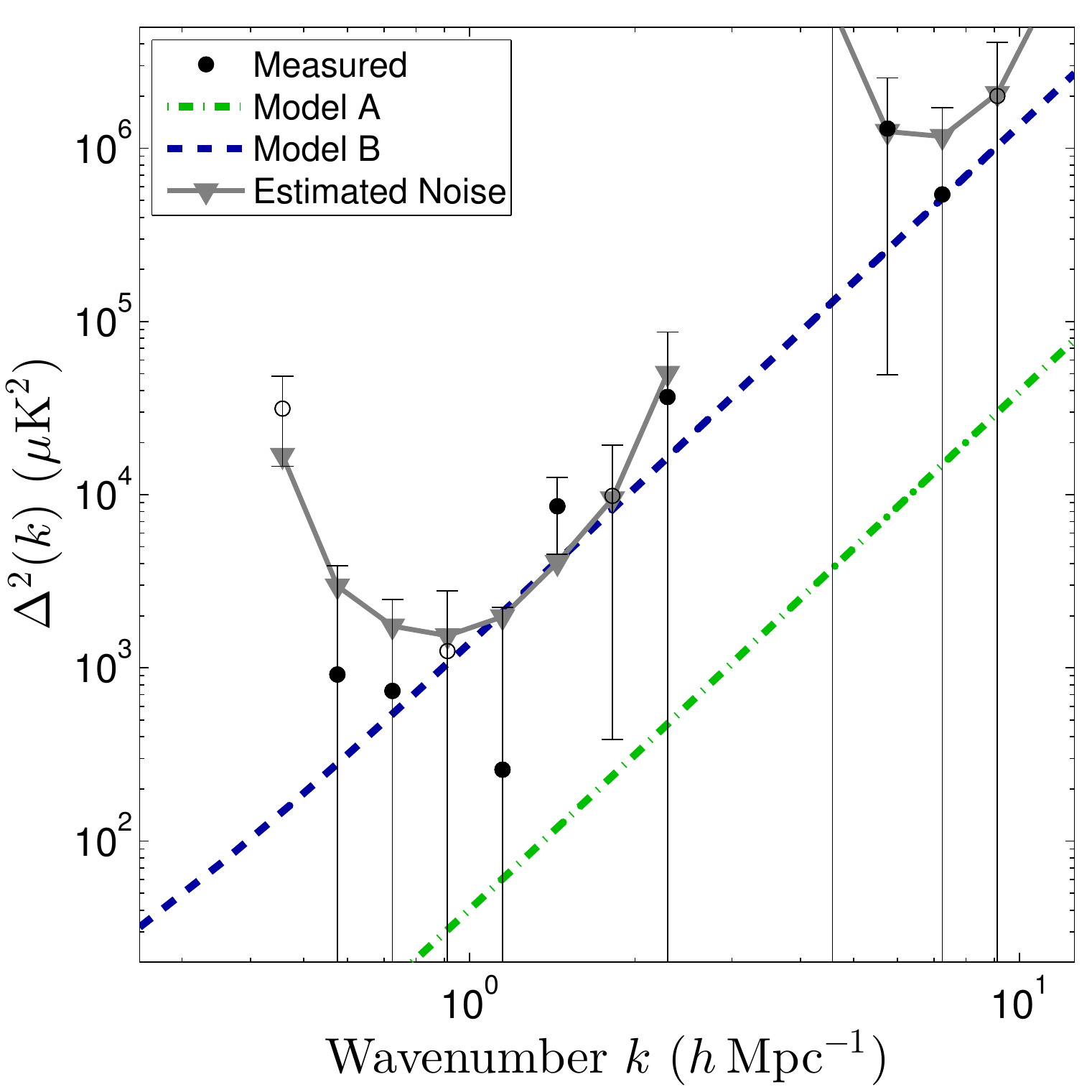}
\end{center}
\caption{\label{fig:sens} \small
Power spectrum of CO intensity from $z=2.3$ to 3.3.  We show two
different models (A and B) for the signal strength based on observed and
theoretical star-formation estimates \cite{2013ApJ...768...15P}.
Sunyaev-Zel'dovich Array results from  Keating et al. (2015) are shown
\cite{2015AAS...22542705K}.
}
\end{figure}

Intensity mapping probes the distribution of lower mass galaxies through
their spatial fluctuations. This signal is identifiable in the variance
($\Delta^2_\mathrm{CO}$) of the spatial+spectral power spectrum, as a
function of the three-dimensional wave-vector ($k$).
The technique is a three-dimensional version of the two-dimensional
techniques used to characterize the cosmic microwave background (CMB)
anisotropy and is identical to the technique that has been developed for
hydrogen epoch of reionization experiments \cite{2004ApJ...615....7M}.
The CO intensity power spectrum consists of two components: a Poisson
component from the random distribution of individual galaxies and a
component from the large-scale clustering of the galaxies imposed by the
process of structure formation.  The transition between these components
occurs at $k \sim 0.3$ h Mpc$^{-1}$, as can be seen in the power-law
break in the models (Fig.~\ref{fig:sens}).  In the current understanding
of the reionization process, the clustered formation of galaxies sets
the distribution of ionizing bubbles, and so sampling these low-$k$
modes is of great interest. 

The ngVLA can make a substantial contribution in this area.  The
proposed frequency range provides sensitivity to multiple transitions of
CO at redshifts from $z\sim1$ to the EoR.  If designed to a maximum
frequency of 115 GHz, 
the $2\rightarrow 1$ transition of CO becomes available at $z>1$,
joining the $1\rightarrow 0$ transition that is accessible, by design,
at z=0. The broad frequency coverage of the array therefore permits
detection of the growth of molecular gas through cross-correlation
between CO transitions starting when the universe was less than half its
current age. Other molecular tracers, such as formaldehyde and fainter
lines, will also be accessible to ngVLA.  Experiments can also be
matched to individual redshift windows from deep optical spectroscopic
surveys, such as those that will be carried out by Subaru, HETDEX and
DESI.  

The primary technical challenge for ngVLA is to provide a wide field of
view and sensitivity to the shortest baselines.  Large dishes and long
baselines will filter out the clustered component of the intensity mapping
signal.  Sensitivity to scales as large as tens of arcminutes is
necessary to detect the clustered component.  This may be feasible with an
array of small, densely-packed dishes at wavelengths $\sim 2$ cm ($z
\sim 6$).  Total power capability for the antennas would also provide
the ability to mosaic large-scales and recover the clustered component.

\subsection{Astrometry}

Most cosmological observables, such as redshift, distance, or flux, are
functions of time:  given enough time or enough precision, these
quantities will be seen to drift due to the Hubble expansion and the
expansion's acceleration.  The rate for real-time cosmological changes
is of order $H_0 \sim 10^{-10}$~yr$^{-1}$, so the necessary precision in
photometry or distance determinations does not exist (although it might
be possible in the coming decades to measure the secular redshift drift
\cite{2012ApJ...761L..26D}).  When expressed as an angular motion,
however, $H_0 \approx 15$~$\mu$as~yr$^{-1}$, which is in the realm of
possibility for current and future telescopes
\cite{2013ApJ...777L..21D}.
 
Correlated proper motions of AGN or quasars serving as test masses can
reveal new ``real-time'' cosmological effects.  Most extragalactic radio
sources show significant intrinsic proper motion at VLBI resolution,
which frustrates proper motion studies of individual objects.
Large-scale correlated proper motions, however, are independent of
internal processes and provide detectable signals at the few-$\mu$as
level, as first demonstrated with the direct detection of the secular
aberration drift caused by the Solar acceleration about the Galactic
Center \cite{2011A&A...529A..91T}.  

A proper motion power spectrum can be used to characterize a vector
field on the sky.  Like the cosmic microwave background (CMB)
fluctuations, which form a scalar power pattern on a sphere and can be
decomposed into spherical harmonic amplitudes for each degree $\ell$,
the proper motion power spectrum can be characterized using {\it vector}
spherical harmonics (VSH), which are the gradient and curl of the scalar
spherical harmonics.  The curl-free VSH resemble electric fields
($E$-modes, or spheroidal modes), and the divergenceless VSH resemble
magnetic fields ($B$-modes, or toroidal modes)
\cite{2012A&A...547A..59M}.  

An ngVLA with VLBA baselines performing astrometric monitoring of 10,000
objects with 10~$\mu$as~yr$^{-1}$ astrometry per object would enable the
global detection of 0.1~$\mu$as~yr$^{-1}$ correlated signals at
5$\sigma$ significance, which is $\sim$0.7\% of $H_0$.  This proper
motion sensitivity would provide the opportunity to detect numerous
observer-induced and cosmological effects:  

$\bullet$ The solar motion with respect to the CMB is 80 AU~yr$^{-1}$,
inducing a distance-dependent $E$-mode dipole with amplitude
$D^{-1}(\rm{Mpc})\times 80$~$\mu$as~yr$^{-1}$.  It may thus be possible
to obtain individual parallax distances to galaxies out to about 8 Mpc
(but peculiar motions complicate this prospect).  

$\bullet$  Anisotropic Hubble expansion would in its simplest form
appear as an $E$-mode quadrupole \cite{2014MNRAS.442L..66D}, and an
expansion vorticity would manifest as a $B$-mode quadrupole.  The ngVLA
could constrain the isotropy of expansion to of order 0.1\% of $H_0$.

$\bullet$  Very long-period gravitational waves will deflect light from
distant objects in a quadrupolar (and higher order) $E$- and $B$-mode
pattern on the sky \cite{2011PhRvD..83b4024B}.  The sensitivity of
astrometry to gravitational waves spans frequencies from
$\sim$1~yr$^{-1}$ to $H_0$ (10 orders of magnitude;
$10^{-18}$--$10^{-8}$ Hz), and the proposed ngVLA astrometric program
could be competitive with pulsar timing (but probes a different and much
larger frequency regime).  

$\bullet$ At higher $\ell$, the real-time growth and recession of large
scale structure will manifest in $E$-modes \cite{2013ApJ...777L..21D}.
Astrometric close pairs embedded in the same large scale structure will
show a convergent proper motion distinct from randomly selected
astrometric pairs of test masses and can in principle measure $H_0$.  

$\bullet$ At $\ell\sim40$, the baryon acoustic oscillation (BAO;
\cite{2005ApJ...633..560E}) scale, the real-time evolution of the BAO
will show a convergent ($E$-mode) signal.  At $z=0.5$, $\theta_{BAO} =
4.5^\circ$, and $d\theta_{BAO}/dt_0 = -1.2$~$\mu$as~yr$^{-1}$.  The
real-time evolution of the BAO measures the ratio $H_0/D_A(z)$ (to first
order).

Several of these real-time effects have not yet been developed
theoretically, so the magnitudes of the anticipated signals are not
well-constrained.  The theory will be addressed prior to the full Gaia
data release and should be mature by 2017.

\section{Technical Requirements for Key Science}

\subsection{Galactic Center Pulsars}

Technical requirements for the Galactic Center pulsar detection and timing are:

\begin{itemize}
\item{Frequency coverage in 10 -- 30 GHz range;}
\item{Ability to produce a phased array beam with a substantial fraction of the collecting
area.  This beam is most readily achieved through a compact array configuration. For
detection, the beam should have a size $\sim 0.5$ arcsec with a large fraction
of the collecting area at 10 GHz, corresponding to $\sim 50\%$ of the collecting area within 
6 km.  This can be mitigated to some degree by tiled searching or multi-beam searching.
For reference, the S2 star has a semi-major axis of 0.12 arcsec.  
For timing, the beam can be more compact.  It is possible
to achieve the necessary beam for timing with an extended array and fast, real-time phasing solutions.}
\item{The phased array beam should have a time resolution of 100 microseconds and total bandwidth
of 10 GHz.}
\end{itemize}

\subsection{Explosive Transients including EM GW Sources}

Technical requirements for transient science are:
\begin{itemize}
\item{Broad frequency coverage for SED characterization;}
\item{Sub-arcsecond beams at all frequencies for source identification and localization;}
\item{Survey speed sufficient to image 100 deg$^2$ at 10 $\mu$Jy rms continuum sensitivity in 12 hours at 10 GHz:  This is nominally achievable with an 18-meter dish;}
\item{Real-time, automated telescope control to respond within minutes to triggers followed by
real-time analysis and delivery of data products.}
\end{itemize}

\subsection{Plasma Physics}

While some plasma physics problems may require very high resolution or simply
point source sensitivity, the problems focused on here require imaging of
large objects with relatively modest spatial resolution but high surface
brightness sensitivity and fidelity. This implies the following:

\begin{itemize}
\item{High surface bright sensitivity at 1-10 arcsec resolution. Implies a
compact core for the array with at least 50\% of the collecting area.}

\item{Good mosaicking over fields of up to at least 30x30 arcmin. Argues for small
dishes, perhaps 12m, with stable, well-known primary beams, to allow the
smallest number of pointings and good restoration of mosaicked images.}

\item{A broad frequency range using as few separate receivers as possible between
10-100 GHz in order to increase the effective sensitivity for spectral shape
measurements.}

\item{Some way to sample spatial frequencies on scales larger than allowed by
the minimum spacing in the main array. This may require a single dish.}

\end{itemize}

 We need good surface brightness imaging, because the
critical features which can discriminate between these models are
small and bright, but also embedded in larger, extended emission
regions.  We need high frequencies because spectra are a key test, and
a good spectral sampling across a broad frequency range (at least a
decade, say) is needed to measure spectra robustly. 

Since the smallest current configuration of the VLA would produce
$\sim 1$ arcsec resolution at 100 GHz, ngVLA must have a very compact 
configuration with enough collecting area to yield better surface
brightness sensitivity than the VLA. Furthermore, the large FOVs
required means that ngVLA must work well as a a mosaicking array,
since the individual telescopes will have much smaller primary beams
than many of the sources of interest, e.g. the Crab, A2256 and
M87. This requirement also argues for smaller array elements, perhaps
12m, in order to obtain adequate pointing for a mosaic and to reduce
the number of pointings needed. Furthermore some total power mode,
perhaps like ALMA or using a larger single dish, to sample the
largest spatial frequencies, is needed. Given the large number of 
astrophysically interesting sources larger than 30
arcsec, this requirement is likely to be important for many areas of 
research with ngVLA. For the difficult case of A2256 assuming a
compact configuration with 3X the VLA collecting area,
spectral index of -0.7 and 12m antennas, the relic could
be imaged in the 10-50 GHz band using a 400 pointing mosaic with
typically 20:1 S/N with 3 arcsec resolution in 25 hours.  

\subsection{Exoplanet Space Weather}

We consider requirements for two goals.

Goal: confirm detections of stellar wind around nearby planet-hosting star, using inferences based on recent papers.  Requirements are:

\begin{itemize}

        \item{RMS sensitivity of 100 nJy at 10 GHz for 5 sigma detection of radio emission from ionized wind with mass loss rate of 100 times solar mass loss rate, low velocity wind, M dwarf at 10 pc.}
        \item{RMS sensitivity of 50 nJy at 10 GHz for 5 sigma detection of radio emission from ionized wind with mass loss rate of 25 times solar mass loss rate, low velocity wind, M dwarf at 5 pc.}
        \item{Instantaneous frequency coverage of 3 GHz at 10 GHz to confirm spectral index expected for optically thick/thin emission.}
\end{itemize}

Goal: diagnose magnetospheric interactions of magnetized close-in planet with host star, for a star with radio luminosity a few times $10^{13}$ erg/s/Hz (at the faint end of current radio detections), at a distance of 10 pc. This is sufficient to encompass the distance of any likely Earth-like planets in their host star's habitable zones. 
Requirements are:

\begin{itemize}
        \item{Instantaneous frequency coverage of 3 GHz or greater to diagnose changing plasma conditions.}
        \item{RMS sensitivity over 5 minutes of 1 microJy (should be equivalent to 100 nJy in 10 hours) over a bandwidth of 500 MHz, sufficient to detect robustly radio emission from a stellar source 10 pc away with $L_r=$a few times $10^{13}$ erg/s/Hz, and determine changing flux levels at the level of 20\% to a significance of several sigma, as a function of frequency and time.}
        \item{Sensitivity to detect circularly polarized emission at the level of 20\% of the total intensity in a timescale of 5 minutes, in a source with a radio luminosity of a few times $10^{13}$  erg/s/Hz at a distance of 10 pc.}
\end{itemize}

\subsection{Astrometry}

Here we discuss technical considerations to achieve optimal VLBI astrometry
for the ngVLA.  The precision of relative astrometry ("in-beam") depends 
on the size of the synthesized beam and the sensitivity on long baselines.
For absolute position determination, the critical parameter is simply the
baseline length, along with delay error.  Therefore, the one stringent 
requirement is that at least 20\% of the collecting area of the array 
should cover long baselines, up to $\sim$5000 km, in both the east-west and 
north-south directions.  For astrometry of compact targets, e.g. masers, 
good uv coverage is not essential, so the long baselines could in principle 
be met with 5-7 large dishes, or phased groups of smaller antennas, that 
would supplement the main central array.  The antennas should cover observing 
frequencies at least up to 22 GHz.

Absolute position determination further depends on the quality of the
geodetic calibration (i.e. residual atmospheric and clock delays) and phase
transfer from astrometric calibrators.  The technical requirements here are 
mild.  The antennas should be able to slew fairly quickly, 
preferably $> 40$ degrees per minute, and the system should allow observing at 
least 1 GHz continuum bandwidth.

Finally, maser astrometry requires spectral observing modes that can cover
1 GHz contiguously with at least 25 kHz spectral resolution, to adequately
sample the narrow line profiles that cover wide velocity ranges.


\end{document}